\documentstyle [aps,multicol,psfig]{revtex} 

\begin{document}

\title{Thermodynamic properties of a small superconducting grain}

\author{M. Schechter$^1$, Y. Imry$^1$, Y. Levinson$^1$, and J. von Delft$^2$} 

\address{$^1$Dept. of Condensed Matter Physics, The 
Weizmann Institute of Science, Rehovot, 76100, Israel
\newline  
	$^2$Physikalisches Institut, Universit\"at Bonn, D-53115 Bonn, Germany}

\date{December 9, 2000}

\maketitle 
  
\begin{abstract} 
The reduced BCS Hamiltonian for a metallic grain with a finite number
of electrons is considered.  The crossover between the ultrasmall
regime, in which the level spacing, $d$, is larger than the bulk
superconducting gap, $\Delta$, and the small regime, where $\Delta
\gtrsim d$, is investigated analytically and numerically.  The
condensation energy, spin magnetization and tunneling peak spectrum
are calculated analytically in the ultrasmall regime, using an
approximation controlled by $1/\ln N$ as small parameter, where $N$ is
the number of interacting electron pairs.  The condensation energy in
this regime is perturbative in the coupling constant $\lambda$, and is
proportional to $d N \lambda^2 = \lambda^2 \omega_D$.  We find that
also in a large regime with $\Delta>d$, in which pairing correlations
are already rather well developed, the perturbative part of the
condensation energy is larger than the singular, BCS, part.  The
condition for the condensation energy to be well approximated by the
BCS result is found to be roughly $\Delta > \sqrt{d \omega_D}$.  We
show how the condensation energy can, in principle, be extracted from
a measurement of the spin magnetization curve, and find a re-entrant
susceptibility at zero temperature as a function of magnetic field,
which can serve as a sensitive probe for the existence of
superconducting correlations in ultrasmall grains.  Numerical results
are presented which suggest that in the large $N$ limit the $1/N$
correction to the BCS result for the condensation energy is larger
than $\Delta$.

\end{abstract} 

\begin{multicols}{2} 

\section{Introduction and Summary of Results} 

In the macroscopic limit, a system described by the reduced BCS
Hamiltonian is well treated by the mean field BCS method \cite{BCS57}.
When the size of a superconducting sample becomes small, two related
questions can be asked: what is the lower size limit for which
superconducting properties are observable, and what is the lower size
limit for the validity of the BCS theory?

In 1959 Anderson \cite{And59} considered the first question, and
argued that ``superconductivity would no longer be possible'' once the
electron spectrum's mean level spacing, $d$, becomes larger than the
bulk superconducting gap, $\Delta$.  ($1/d ={\cal N}(0)$, the density
of states per spin species near the Fermi energy, hence $d \propto
1/{\rm Volume}$.)  This statement sets a lower limit for the size
above which a grain still exhibits superconducting properties, but at
the same time states that such a grain can well be much smaller than
the superconducting coherence length.  Superconductors in the regime
where the level spacing is comparable to the gap energy have been
studied for many years both theoretically (e.g., Ref.~\cite{MSD72})
and experimentally (e.g., Ref.~\cite{GZ68}, see also review by
Perenboom {\it et. al.} \cite{PWM81}).

Recently, Ralph, Black and Tinkham (RBT) performed measurements on
single superconducting nm-scale grains in the regimes of $\Delta
\gtrsim d$ and $\Delta \lesssim d$
\cite{BRT96}. These experiments, and the 
considerable amount of theoretical work they initiated
\cite{SA96,DZGT96,BDRT97,ML97,AA97,BD98,BH98,BD99,DS99,DB00,DS00},
found various properties indicative of strong superconducting pairing
correlations in grains with $\Delta \gtrsim d$ (to be called ``small
grains''), but not in grains with $\Delta < d$ (to be called
``ultrasmall grains''), thus supporting Anderson's criterion.  These
properties include (i) a parity dependent gap in the excitation
spectrum (the gap exists only for grains with an even number of
electrons), which is driven to zero by magnetic field
\cite{BRT96,DZGT96,BDRT97,BD99}; (ii) a difference of order
$\Delta$ in the ground state energies of even and odd grains
\cite{ML97,BH98}; 
and (iii) a first order paramagnetic transition induced by a magnetic
field \cite{BDRT97,BD99}.

Though {\em ultra}small grains with $\Delta < d$ do not have as
strongly developed signatures of pairing correlations as the small
grains with $\Delta \gtrsim d$ mentioned above, pairing correlations
nevertheless do exist in such grains, albeit in the form of weaker
fluctuations, and they can affect various physical quantities. For
example, Di Lorenzo {\it et. al.}
\cite{LFH+00} find that pairing correlations affect the temperature 
dependence of the spin susceptibility of grains also in the ultrasmall
regime. 

The crossover regime 
between small and ultrasmall grains has also been studied in 
some detail numerically, using a simple reduced BCS model
with a discrete set of single-particle levels. 
\cite{BD98,BH98,DS99,DS00}.
In particular, it was found that the condensation energy, $E_{cond}$
(i.e. the energy gain of the exact ground
state relative to the uncorrelated Fermi ground state), 
smoothly crosses over from being extensive 
(proportional to the size of the system) 
for $\Delta>d$ to being intensive for $\Delta<d$. 

One of the goals of the present paper is to obtain further insights
into the crossover from the ultrasmall regime, which can be treated
perturbatively in the dimensionless coupling ($\lambda$) of the said
reduced BCS model, to the small regime, which can not.  Our point of
departure is an exact solution, due to Richardson \cite{Ric63,RS64},
of the reduced BCS model of present interest.  By analyzing
Richardson's solution both analytically and numerically in the
crossover regime, we elucidate in detail when and how perturbation
theory in $\lambda$ breaks down, how the answer depends on the system
size, and how the standard BCS results are recovered in the bulk limit
$d \ll \Delta$.

The bulk regime is of course well-known to require a nonperturbative
treatment; indeed, the BCS result for the condensation energy,
\begin{equation}
  \label{eq:EcondBCS}
  E^{BCS}_{cond} = \Delta^2/(2d) \; ,
\end{equation}
is not analytical in $\lambda$ as $\lambda \to 0$, since the bulk 
gap is given by
\begin{equation}
  \label{eq:bulkgap}
  \Delta (\lambda) = \omega_D / \sinh (1/\lambda) \qquad \mbox{[}\simeq
 2 \omega_D e^{- 1/ \lambda} \quad 
\mbox{for} \; \; \lambda \ll 1 \mbox{]}\;,
\end{equation}
where $\omega_D = N d$ is the bandwidth about the
Fermi energy within which the pairing interaction acts 
(typically the Debye frequency). This nonanalyticity arises
because BCS consider the thermodynamic limit of infinite
system size ($N \to \infty$, $d \to 0$, at fixed $\omega_D$). 

We shall argue that
if instead one considers a system with a finite number
of pairs, say $N$,
the condensation energy $E_{cond} (\lambda)$ 
is an analytical function about $\lambda = 0$,
with a finite radius of convergence given approximately 
by $ \lambda^\ast = 1/ \ln N$. For 
$\lambda < \lambda^\ast (1- \lambda^\ast)$, 
corresponding to $\Delta < d$ 
(by Eq.~(\ref{eq:bulkgap}) \cite{energies}). 
$E_{cond} (\lambda)$ is found to be well approximated by
the perturbative result,
\begin{equation}
  \label{eq:Econd-approximate_int}
  E^{pert}_{cond} (\lambda)
  = \ln{2} \cdot \lambda ^2 \omega_D \; . 
\end{equation}
On the other hand, the BCS mean-field
result, $E_{cond}^{BCS}$ of (\ref{eq:EcondBCS}),
is found to become reliable only
for $\lambda > 2 \lambda^\ast$, corresponding
to roughly $\Delta > \sqrt {\omega_D d}$. 
Thus, we identify a substantial intermediate regime, 
\begin{equation}
  \label{eq:regimeII}
\lambda^\ast < \lambda < 2 \lambda^{\ast} \; , 
\qquad \mbox{i.e.} \qquad 
   d < \Delta <  \sqrt {\omega_D d} \; ,
\end{equation}
in which neither the perturbative result nor the BCS mean-field result
adequately reproduces $E_{cond}$ (though, roughly speaking, the sum
$E^{pert}_{cond} + E^{BCS}_{cond}$ does).

The existence of this intermediate regime, which to the best of our
knowledge has not been identified before, implies that the regime of
validity of the BCS mean-field approach for calculating $E_{cond}$ is
significantly smaller than realized hitherto: the crossover level
spacing ($d> \Delta^2/\omega_D)$ beyond which it becomes inadequate is
considerably smaller than the scale $(d >\Delta)$ beyond which the BCS
approach formally breaks down (in the sense of yielding no non-trivial
solution to the self-consistency equation
\cite{DZGT96}), and up to which strong signatures for pairing correlations can 
still be observed, as mentioned above.

We are also able to pinpoint the reason for the failure of the BCS
approach in the intermediate regime (\ref{eq:regimeII}): we shall show
in detail that $E_{cond}^{BCS}$ incorporates only contributions to
$E_{cond}$ from the strongly pair-correlated, ``condensed'' levels
within $\Delta$ of the Fermi energy $E_F$, but neglects contributions
from all the remaining, ``weakly pair-fluctuating'' levels, which
extend to a distance $\omega_D$ from $E_F$.  Although the latter
levels are so weakly correlated that their contribution can be
calculated perturbatively, essentially yielding $E_{cond}^{pert}$,
this contribution turns out to be larger than $E_{cond}^{BCS}$ as long
as $\Delta < \lambda \sqrt{\omega_D d}$, and is not negligible
compared to the $E_{cond}^{BCS}$ in the whole intermediate regime
(\ref{eq:regimeII}).  [Note though, that $E^{pert}_{cond}$ would
largely cancel out when one considers energy differences between
eigenstates that differ only in the specific placement of a small
number of electrons in levels near $E_F$. An example would be the
ground state energy difference between an even and odd superconducting
grain, for which the BCS approach would be adequate in the
intermediate regime.] 

It should be mentioned here that the question of how to recover the BCS
gap equation from Richardson's exact solution has been solved
by Richardson himself \cite{Ric77}, by effectively doing a $1/N$ 
expansion around the bulk, thermodynamic limit. Our work differs 
from his in that we do an expansion in $\lambda$ around the ultrasmall
limit for a  system of finite size, with $\lambda < 1/\ln N$ as small
parameter. 

Using the insights gained from our studies of the condensation energy,
we also calculate various other thermodynamic properties of ultrasmall
grains at zero temperature, using a controlled analytical
approximation with $\lambda < 1/ \ln N$ as small parameter.
Specifically, we calculate the spin magnetization and susceptibility
curves and tunneling peak spectrum of ultrasmall grains, and find that
pairing correlations have their signature in all the above physical
quantities, even in the regime $\lambda < \lambda^\ast$ where pairing
correlations are weakest.

The condensation energy can, in principle, be measured by integrating
the spin magnetization as a function of magnetic field (H) and
comparing it to the linear curve of a normal grain.  In fact, as we
discuss in Sec.~\ref{secmagnetization}, since the energy levels in the
grain are not equally (or systematically) spaced, one needs to do the
measurement on an ensemble of grains.  Calculating the spin
susceptibility of an ultrasmall grain, we find that for $H \gg d /
\mu_B$, pairing fluctuations of levels far away from $E_F$ result in a
correction of order $\lambda^2 d / \mu_B H$ to the normal
susceptibility.  Interestingly, this correction persists for all
fields $H < \omega_D/\mu_B$, i.e.\ well beyond the
Clogston-Chandrashekar field $\mu_B H_{CC} =\Delta / \sqrt 2$
\cite{Clo62}, at which, for bulk systems, a first order
transition occurs from the superconducting ground state to a
paramagnetic ground state.  (Only for $H > \omega_D/\mu_B$, the grain
becomes effectively ``normal'', since then all the levels within
$\omega_D$ from the Fermi energy become unpaired.)  The correction {
to} the spin susceptibility results in a re-entrant behavior of the
differential susceptibility as a function of magnetic field, which
could possibly serve as a sensitive probe to detect superconducting
correlations in ultrasmall grains \cite{LFH+00}.  { (The consequences
of pairing correlations in the regime $H > H_{CC}$ have also been
studied by Aleiner and Altshuler \cite{AA97}, who found an anomaly in
the tunneling density of states).}

Similarly, we argue below that in ultrasmall superconducting grains,
pairing fluctuations involving levels far away from $E_F$ are
sufficiently strong that they also leave their mark in the specific
heat (even for $T { \gg} T_c$), and in the tunneling peak spectrum. 

All our calculations are done for grains with an even number of
electrons.  
The results for grains with an
odd number of electrons are similar in the ultrasmall regime, 
and will be  discussed shortly for each calculated quantity.

The paper is arranged as follows: 

In Sec.~\ref{secusenergy} we  calculate the condensation energy of 
an ultrasmall superconducting grain in the regime $\Delta < d$, 
and also analyze the intermediate regime  of Eq.~(\ref{eq:regimeII})
for larger grains. 
In Sec.~\ref{secmagnetization} the spin magnetization of ultrasmall
grains as function of magnetic field is calculated. It is shown that
the condensation energy is given by integrating the magnetization from
$H=0$ to $\omega_D/\mu_B$. 
In Sec.~\ref{secsusceptibility} we calculate the differential spin 
susceptibility of ultrasmall grains as a function of magnetic field, 
and find that it exhibits a re-entrant behavior.  
In Sec.~\ref{sectunneling} the tunneling peak spectrum of an ultrasmall 
superconducting grain is calculated.
In Sec.~\ref{secRichardson} we present numerical results for the 
contribution of the ``condensed'' and ``fluctuating'' levels to the 
condensation energy.

The technical aspects of our calculations are collected
in three appendices. 
In App.~\ref{appaccuracy} a detailed derivation of the accuracy of the 
condensation energy approximation is given.
In App.~\ref{appfactorial} the functional behavior of the prefactors of 
the series expansion of the approximate condensation energy is analyzed.
In App.~\ref{appseries} the series expansion of the exact condensation 
energy is discussed.

\section{Condensation energy of an ultrasmall grain}
\label{secusenergy} 

\subsection{Richardson's Equations}

We consider the reduced BCS Hamiltonian: 

\begin{equation}
\hat{H} = \sum_{j,\sigma=\pm} \epsilon_j c^\dagger_{j\sigma} c_{j\sigma} 
- \lambda d \sum^I_{i,j}  c^\dagger_{i+} c^\dagger_{i-} c_{j-} c_{j+} .
\label{Hamiltonian}
\end{equation}
for a grain with a given, finite number of electrons $\bar{N}$. 
The first term is the kinetic term, which we will refer to as $\hat{H_0}$, 
and the second term is the interaction Hamiltonian, denoted $\hat H_I$. 
The sum in $\hat H_I$ is over all the levels inside the range 
$E_F - \omega_D < \epsilon < E_F + \omega_D$, which we designate by $I$. 
The Hamiltonian~(\ref{Hamiltonian}) is the usual BCS Hamiltonian used 
when discussing superconducting grains 
\cite{DZGT96,BDRT97,ML97,AA97,BD98,BH98,BD99,DS99,DB00,DS00,LFH+00} 
and its validity is discussed 
in, e.g., Refs.~\cite{AA97,Aga99,DR00}. 
(In particular, for the model to be valid the grain's 
dimensionless conductance, 
$g$, must be much larger than one). 
In all cases discussed below we  
consider states in which all levels below $E_F - \omega_D$ are doubly 
occupied, while all levels above $E_F + \omega_D$ are empty. Since the 
dynamics  of electrons occupying levels outside the range $I$ and their 
contribution to the total energy are trivially given by $\hat{H_0}$, we 
will not consider them henceforth.

Richardson and Sherman
\cite{Ric63,RS64} showed that this Hamiltonian, with a finite number
of electrons, can be solved exactly.  They define for each single
particle eigenstate of $\hat{H_0}$ the operator $\xi_j =
c^\dagger_{j+} c_{j+} - c^\dagger_{j-} c_{j-}$. 
This operator, for any
$j$, is a constant of motion of the Hamiltonian (\ref{Hamiltonian}),
and takes the value $\pm 1$ if the level is singly occupied, and 0
otherwise.  The many body eigenstates of (\ref{Hamiltonian}) can
therefore be classified into different subspaces according to their
value of the $\xi_j$'s, i.e., according to the configuration of levels
within $I$ which are occupied by one electron only.  The many body
eigenstates and the eigenenergies of (\ref{Hamiltonian}) are then
found separately \cite{Ric63,RS64} for each of the above subspaces.
 
The electrons in the singly occupied levels are not scattered to other
levels by the interaction term, and the singly occupied levels are
``blocked'' to pair scattering, and we therefore designate them by
$B$.  The dynamics of the singly occupied levels is
also trivially given by $\hat{H_0}$.  Therefore, for each set $B$ one
has to solve the reduced Hamiltonian:

\begin{equation}
\hat{H} = \sum^U_{j} 2 \epsilon_j b^\dagger_{j} b_{j} 
- \lambda d \sum^U_{i,j}  b^{\dagger}_{j} b_{j} \; .
\label{HamiltonianU}
\end{equation}
{Here $b^{\dagger}_j = c^{\dagger}_{j+} c^{\dagger}_{j-}$
creates a {\em pair} of electrons
in level $j$,
and $U$ is} the set of paired levels within $I$, i.e. the set of all 
levels that belong to $I$ but not to $B$
(the notation, in general, follows Ref.~\cite{DR00}).
Below, sums over levels labeled by $j$ 
are to be understood as sums over levels within $U$.

Once the configuration of unpaired electrons is given, 
Richardson and Sherman (\cite{RS64}, see \cite{DB00} for a review) 
show that the eigenstates of the system 
are given by: 

\begin{eqnarray}
|\alpha \rangle = \prod_{i \in B} c^\dagger_{i \sigma_i} | \Psi_k \rangle \, , 
&  \; \; \; \;   
| \Psi_k \rangle = C \prod_{\nu = 1}^k B^\dagger_\nu|0 \rangle \, ,  \nonumber 
\\ 
B^\dagger_\nu = \sum_j \frac{b^\dagger_j}{2 \epsilon_j - E_\nu} , 
\end{eqnarray} 
where $2k$ is the number of electrons occupying the unblocked levels, 
and $|0 \rangle$ 
is the state with all the levels below $E_F - \omega_D$ fully occupied and 
all the levels above $E_F - \omega_D$ empty (in our model $|0 \rangle$ 
is the vacuum state). 
The energy parameters $E_\nu$ 
(with $\nu = 1, \dots , k$) are the solutions of a set of $k$ coupled 
nonlinear equations, the $\nu$'th equation of which is given by: 
\begin{equation}
\frac{1}{\lambda d} +  \sum_{\mu = 1 (\neq \nu)}^k  
\frac{2}{E_\mu - E_\nu} - \sum_j 
\frac{1}{2 \epsilon_j - E_\nu} = 0 .
\label{Richeq}
\end{equation} 
The total energy of the system is given by \cite{Ric63,RS64}: 
\begin{equation}
E = \sum_j^B \epsilon_j + \sum_{\nu=1}^k E_\nu .
\label{energyeqapp}
\end{equation}
Since the ground state of a grain with an even number
  of electrons does not contain any singly-occupied levels
(i.e.\ $U = I$), the even
  ground state energy is simply $\, E_{g.s.}=\sum_{\nu= 1}^k E_\nu$.
Its $\lambda \to 0$ limit is
 $E_{g.s.}(\lambda= 0) = \sum_{\nu= 1}^k 2
  \epsilon_\nu$, where $\{2 \epsilon_\nu, \nu = 1, \dots, k\}$
is the set of the $k$ lowest-lying
  single-pair energies. [This is consistent with
the observation, following from 
 Eq.~(\ref{Richeq}), that in the limit $\lambda \to 0$
  the set of $E_\nu$'s reduces to a set of $k$ single-pair
  energies $2 \epsilon_j$, which, for the ground
state, must have the lowest total energy possible.]
Consequently, 
the interaction energy of the even ground state, $E_{int}
  (\lambda)$, defined to be the reduction of the exact ground state energy
  as the interaction is turned on from zero to some finite $\lambda$,
 can be written as 
\begin{equation}
  \label{eq:Eint}
  E_{int} (\lambda) \equiv E_{g.s.} (0) - E_{g.s.} (\lambda) 
  = \sum \delta E_\nu \; , 
\end{equation}
where we introduced
the energy differences $\delta E_\nu \equiv 2 \varepsilon_\nu -
E_\nu$. A closely related quantity is
the condensation energy, $E_{cond} (\lambda)$, defined
to be the energy gain of the exact even ground state relative
to the uncorrelated Fermi ground state: 
\begin{eqnarray}
\label{eq:definecond-energy}
  E_{cond} (\lambda) & \equiv & E_{F.g.s} (\lambda)  - E_{g.s}(\lambda) 
\\ & = & \sum_{\nu = 1}^k (2 \epsilon_\nu - \lambda -   E_\nu ) \; 
\label{eq:Econd-Eint}
             = E_{int} (\lambda) - k \lambda d \; . 
\end{eqnarray}
The $\lambda$-contribution in the first sum in 
Eq.~(\ref{eq:Econd-Eint}) 
is the Hartree self-energy of level
$j$ in the Fermi ground state.

\subsection{Perturbative results for $E_{cond}$ and $E_{int}$}

Let us now consider the case that the set $I$ of
interacting levels consists of $2 N$ equally spaced energy levels between 
$E_F - \omega_D$ and $E_F + \omega_D$, occupied by $2 N$ electrons,
so that $k = N$. 
Measuring the single-particle energies w.r.t. to 
the bottom of the interacting band, 
we thus take $\epsilon_j = j d$, where $j = 1, \dots 2N$ and 
$d = \omega_D / N$.
(Note that $N \neq \bar{N}$ 
the total number of electrons in the grain,  
which is of order $2 E_F / d$, not $2 \omega_D/d$). 

Using Eq.~(\ref{Richeq}), the energy differences $\delta E_\nu$
occurring in Eq.~(\ref{eq:Eint}) can be rewritten as  
\begin{equation}
\delta E_\nu \equiv 2 \epsilon_\nu - E_\nu = 
\frac{\lambda d}{1 - \lambda \, a_\nu} \; , 
\label{delEnu}
\end{equation} 
where 

\begin{equation}
a_\nu = d \left( \sum_{j=1(\neq \nu)}^{2N} \frac{1}{2 \epsilon_j - E_\nu} - 
\sum_{\mu=1(\neq \nu)}^N \frac{2}{E_\mu - E_\nu} \right). 
\label{anu}
\end{equation}
For small $\lambda$, it is natural to approximate
$\delta E_\nu$ by
\begin{equation}
\delta E_\nu^0 \equiv \lambda_\nu d, \qquad
\mbox{where} \quad \lambda_\nu \equiv
\frac{\lambda }{1 - \lambda \, a_\nu^0} ,
\label{delEnuapprox}
\end{equation} 
and  $a_\nu^0 \equiv a_\nu (\lambda=0)$ is given by
\begin{equation}
a_\nu^0 =  \sum_{j=1(\neq \nu)}^{2N} \frac{1}{2 j - 2 \nu} - 
\sum_{\mu=1(\neq \nu)}^N \frac{2}{2 \mu - 2 \nu}.
\label{anu0}
\end{equation}
The accuracy of this approximation
is studied in App.~\ref{appaccuracy} (by
deriving an expression for  
$\delta a_\nu =  a_\nu - a_\nu^0$), 
where we find that the
relative error in $\delta E_\nu$
depends on both $\lambda$ and $N$: Specifically, 
we find that for all $\nu$, 
\begin{mathletters}
\label{relativelnall}
\begin{eqnarray}
&&\delta E_\nu / \delta E_\nu^0 = 1 + {\cal O} (1/(\ln{N})^2) \, 
\quad \mbox{for} \quad \lambda < 1/(2 \ln{N}) \,  ,
\label{relativeln} \\
&&\delta E_\nu / \delta E_\nu^0  =  1 + {\cal O} (1/c^2)  \, 
\quad \mbox{for} \qquad \qquad \nonumber \\
&& \; \; \; \; \; \quad \qquad \qquad  1/(2 \ln{N}) < \lambda 
< 1/\ln{N} - c/(\ln{N})^2 \, , 
\label{relativeln-largerlambda}
\end{eqnarray}
\end{mathletters}
for any $c>1$. Note that Eq.~(\ref{relativeln-largerlambda})
implies the emergence of a 
second scale near $\lambda=1/\ln{N}$, namely $1/(\ln{N})^2$.
 
To the accuracy given by Eqs.~(\ref{relativelnall}),
the interaction and condensation energies 
can be approximated by
\begin{equation}
E_{int}^0 = E_{cond}^0 + N \lambda d 
\simeq \sum_{\nu = 1}^N \delta E^0_\nu
= 
\sum_{\nu=1}^N \lambda_\nu d \, ,
\label{condmatveev}
\end{equation}
where $\lambda_\nu$ is given in Eq.~(\ref{delEnuapprox}). 
This result coincides with that obtained by Matveev and Larkin
(Eq. (17) of \cite{ML97}); moreover, our approach allows us
to give  a controlled estimate of the error introduced by
this approximation, both for Eq.~(\ref{condmatveev}) and for our explicit 
calculation of $E_{int}$ and $E_{cond}$ in App.~\ref{appfactorial}. 
Interestingly, Eq.~(\ref{condmatveev}) can be interpreted
as a  sum over the Hartree selfenergies $\lambda_\nu d$ of the lowest $N$
levels, each of which is evaluated 
using its own, level-specific ``renormalized coupling constant''
 $\lambda_\nu$ (thus motivating our choice of notation).
The emergence of such renormalized coupling constants has
been noted before \cite{AGD63}, in particular by Matveev and Larkin \cite{ML97}
and Berger and Halperin \cite{BH98}. Matveev
and Larkin, for example, were concerned with
perturbatively calculating a certain parity parameter that was essentially
equal to $\lambda_N d/2$, and found
\begin{equation}
{\lambda^{ML}_N} \simeq 
\frac{\lambda}{1 - \lambda \ln ( \omega_D/ d)},
\label{renormalization}
\end{equation}
in agreement with our result 
[Eq.~(\ref{delEnuapprox})] for $\lambda_N$ (see 
Eq.~(\ref{anu0approxi}) and the statement following it). 

Now, calculating the interaction or condensation energies is
considerably more involved than calculating the parity parameter of
Matveev and Larkin, since, in contrast to their calculation, not only
one but all $N$ renormalized couplings $\lambda_\nu$ enter in
Eq.~(\ref{condmatveev}) for $E_{int}^0$ or $E_{cond}^0$.  This is a
major complication, since their $\nu$-dependence turns out to be
sufficiently important that it { is} not possible to replace all
$\lambda_\nu$ by a single ``effective coupling constant''.

Nevertheless, progress can be made
by expanding $E_{int}^0$ or $E_{cond}^0$ in powers of $\lambda$
and analyzing the convergence properties of the resulting
series. This is done in  App.~\ref{appfactorial} 
[for $E_{int}^0$, but here we shall give the results
for $E_{cond}^0$, which is slightly more convenient,
since it lacks the Hartree term]. It is
found that the convergence radius of the power series
for $E_{cond}^0 (\lambda)$ is 
\begin{equation}
  \label{eq:lambdastar}
    \lambda^\ast = 1/ \ln N \; . \qquad
\end{equation}
The regime of analyticity, 
\begin{equation}
  \label{eq:convergencecondition}
  \lambda < \lambda^\ast,  \qquad \mbox{i.e.} \qquad
  \Delta < d \; ,
\end{equation}
[by Eq.~(\ref{eq:bulkgap})], will be 
called {``regime I''} below. 
Within regime I, we obtain an analytical expression for 
$E^0_{cond} (\lambda)$ as a series in $\lambda$. 
We find (see App.~\ref{appfactorial}) that the series for $E_{cond}^0$ 
does not have one parameter which describes the ratio between consecutive 
terms in the series. Denoting the $m$'th term in the power series as 
$E^{0(m)}_{cond}$, we show that 
the low powers fulfill the relation: 
$E^{0(m+1)}_{cond}/E^{0(m)}_{cond} \simeq  m \cdot \lambda$ while the high 
powers fulfill the relation: 
$E^{0(m+1)}_{cond}/E^{0(m)}_{cond} \simeq \lambda \cdot \ln{N}$. 
This results in having two separate scales in $\lambda$. 
While the high powers dictate the convergence radius of the series to be 
$\lambda^\ast$, their contribution is large only for 
$\lambda \gtrsim \lambda^\ast (1 - \lambda^\ast)$
(see App.~\ref{appfactorial}), introducing 
the aforementioned second
scale of $1/(\ln{N})^2$ near $\lambda=1/\ln{N}$. 
As a result, for $\lambda < \lambda^\ast (1 - \lambda^\ast)$ 
(i.e.  in most of regime I), $E^0_{cond}$  
is well approximated 
by the contribution of the low powers,
which turn out to correspond simply to the 
second order perturbative result 
(up to a relative correction of $1/\ln{N}$, see 
App.~\ref{appfactorial}): 
\begin{eqnarray}
  \label{eq:Econd-approximate}
  E^0_{cond} \simeq E^{pert}_{cond} (\lambda)
  = \ln{2} \cdot \lambda ^2 \omega_D [1 + {\cal O}(1/\ln{N})] \nonumber \\ 
  \qquad \mbox{for} \quad \lambda < \lambda^\ast (1 - \lambda^\ast) \; .
\end{eqnarray}
This is illustrated in Fig. \ref{figcondensation1024}. 
Intuitively speaking,
this contribution can be attributed to pairing fluctuations 
involving {\em all} the levels in the range $E_F -
\omega_D < \epsilon < E_F + \omega_D$. 

\subsection{Analysis of the intermediate regime $\lambda^\ast < \lambda < 2
\lambda^\ast$} 

Although we are not able to extend the analytical calculation to the
regime of $\lambda > \lambda^\ast$ (i.e. $\Delta > d$), we are able to
draw some conclusions about the value of the condensation energy in
the latter regime: 

First, we note that the perturbative result
(\ref{eq:Econd-approximate})
 for the condensation energy at $\lambda=\lambda^\ast$ 
  is larger than the BCS mean field result (\ref{eq:EcondBCS}) 
as function of $\lambda$, i.e. $E^{pert}_{cond}(\lambda^\ast)  \gg
 E_{cond}^{BCS}(\lambda) $, as long as $\Delta < \lambda \sqrt{\omega_D
 d}$. In this regime  $E_{cond}^{BCS}(\lambda) $ is thus also
  much smaller than the actual condensation energy $E_{cond}(\lambda)$ (since,
 assuming monotonicity of $E_{cond}(\lambda)$ as function of $\lambda$, we
 have $E_{cond}(\lambda) > 
E_{cond}^{pert}(\lambda^\ast) $ for $\lambda > \lambda^\ast$). 
In terms of $\lambda$ and 
$N$ the condition is 
$\ln{[\ln{2} \cdot (\lambda^\ast)^2/2]} + 2/\lambda > \ln{N}$, which, 
for large $N$, is roughly 
$\lambda < 2 \lambda^\ast$. 
[Note that the
exponential dependence of $\Delta$ on $\lambda$
causes a relatively small change in the condition for 
$\lambda$ ($< \lambda^\ast$ versus
$< 2 \lambda^\ast $) to translate into a parametric change in
the condition for $\Delta$ ($< d$ versus $\lambda \sqrt{\omega_D d}$)].
The $E_{cond}^{pert} $ contribution 
 in Eq.~(\ref{eq:fullEcond})
becomes significantly smaller 
(by a factor $\lambda^2$) than
the  $E_{cond}^{BCS}  $
contribution 
only for $\Delta > \sqrt {\omega_D d}$. Thus, we identify
an intermediate regime,

\setlength{\unitlength}{3in} 

\begin{figure}
       \narrowtext 
        \begin{center}  
        \begin{picture}(1.05,0.649)
        \put(-0.018,-0.252){\psfig{figure=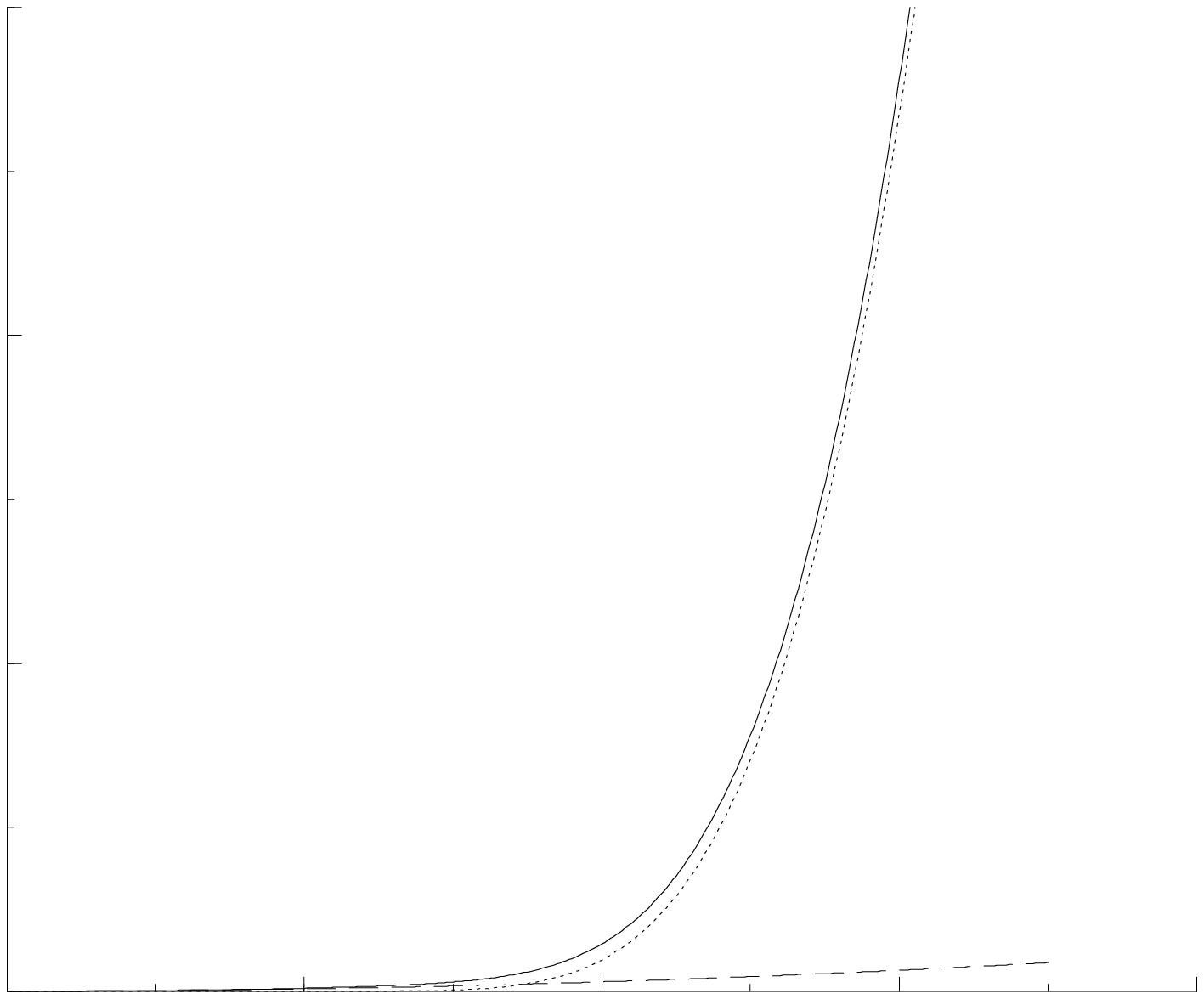,width=3.375in}}
        \put(0.183,0.22){\makebox(0,0)[lb]{\psfig{figure=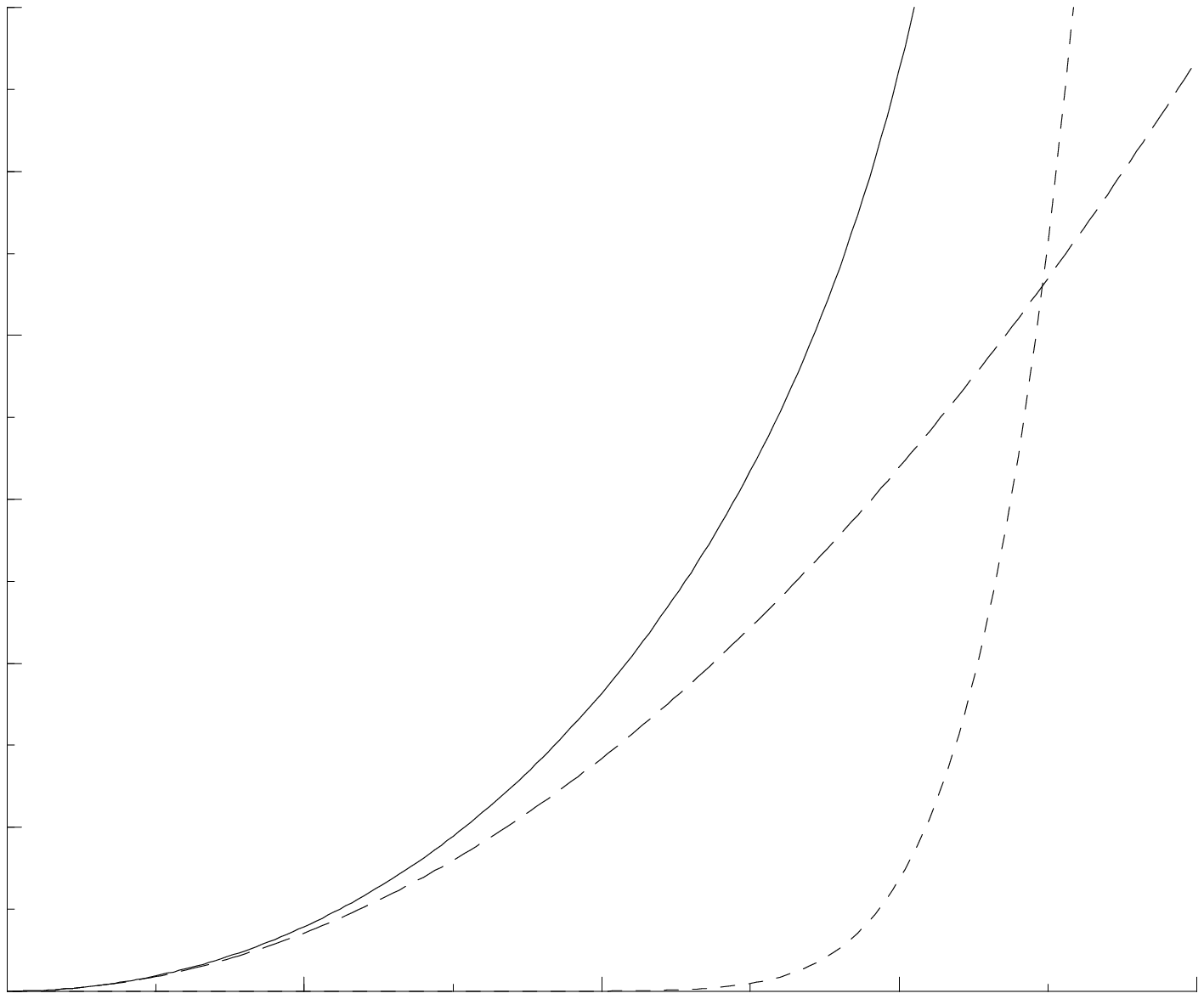,width=1.35in}}} 
        \put(0.073, 0.55){\makebox(0,0)[r]{{$E_{cond}/d$}}}
        \put(0.33, -0.23){\makebox(0,0)[t]{0.1}}
        \put(0.44, -0.27){\makebox(0,0)[t]{$1/\ln{N}$}}
        \put(0.44, -0.22){\makebox(0,0)[t]{$\uparrow$}}
        \put(0.58, -0.23){\makebox(0,0)[t]{0.2}}
        \put(0.83, -0.23){\makebox(0,0)[t]{0.3}}
        \put(1.07, -0.23){\makebox(0,0)[t]{0.4}}
        \put(0.08, 0.065){\makebox(0,0)[r]{1000}} 
        \put(0.08, 0.34){\makebox(0,0)[r]{2000}} 
        \put(0.08, 0.61){\makebox(0,0)[r]{3000}} 
        \put(0.2, 0.415){\makebox(0,0)[r]{15}} 
        \put(0.2, 0.58){\makebox(0,0)[r]{30}} 
        \put(0.41, 0.21){\makebox(0,0)[t]{0.1}}
        \put(0.503, 0.18){\makebox(0,0)[t]{$1/\ln{N}$}}
        \put(0.503, 0.23){\makebox(0,0)[t]{$\uparrow$}}
        \put(0.62, 0.21){\makebox(0,0)[t]{0.2}}
        \put(1.02,-0.26){\makebox(0,0)[t]{$\lambda$}}
        \end{picture} 
        \end{center}
        \vspace{2.5 cm}
        \caption{The condensation 
	energy of a grain with $N=1024$, in units 
        of level spacing, is plotted as a 
        function of $\lambda$. The solid line is the numerical solution 
        of the exact Richardson equations. The 
        dashed line is the second order approximation. The dotted line is 
        the BCS approximation. The BCS approximation is good for 
        $\lambda \gg \lambda^\ast \equiv 1/\ln{N}$. 
	In the inset the same graph is given for a 
        small range of $\lambda$ and a much smaller range for $E_{cond}$. 
        The value at which the perturbative term equals the BCS term tends 
        asymptotically to $2/\ln{N}$ (see text),
       but here it is somewhat smaller
        since $N$ is not very large.}
\label{figcondensation1024}
\end{figure}

\begin{equation}
  \label{eq:regimeIIpa}
  d < \Delta < \sqrt {\omega_D d} \; ,
\qquad \mbox{i.e.} \qquad 
\lambda^\ast < \lambda < 2 \lambda^{\ast} \; 
\end{equation}
[by Eq.~(\ref{eq:bulkgap})], to be called ``regime II'', in which the BCS
mean-field approach is severely inadequate for calculating $E_{cond}$, but
which, according to the three properties mentioned in the introduction,
nevertheless already features strongly-developed pairing correlations.  In
other words, the condition 
for the adequacy of the BCS mean-field
approximation ($\lambda > 2 \lambda^{\ast}$, ``regime III'') is more
restrictive than the condition for the existence of strongly-developed pairing
correlations ($\lambda > \lambda^{\ast}$).  Importantly, this also means that
the BCS mean-field approach becomes inadequate already for much smaller level
spacings, { $d \approx \Delta^2/ \omega_D$,} than those at which it
formally breaks down (in the sense of yielding no non-trivial solution to the
self-consistency equation), which occurs for $d \gtrsim \Delta$.

The inadequacy of the BCS approximation in regime II
stems from the abundance of ``fluctuating'' levels compared 
to ``condensed'' levels. Each ``condensed'' 
level within a range $\Delta$ from the 
Fermi energy contributes approximately $\Delta/2$ to the 
condensation energy, 
and having $\Delta/d$ such levels gives the 
BCS term $\Delta^2/2d$. Though each 
``fluctuating'' level 
outside this range contributes only an amount of order
$(d \lambda)^2 /d$ 
to the condensation energy, 
there are $\omega_D/d$ such levels, and for 
$\Delta < \lambda \sqrt{\omega_D d}$ 
the total contribution $\lambda^2 \omega_D$ 
of all fluctuating levels is larger than $\Delta^2/2d$. 
This sets an energy scale, $\sqrt{\omega_D d}$, which 
$\Delta$ has to exceed before
the BCS approximation becomes reliable. 

The above interpretation of the relative contributions
of ``condensed'' and ``fluctuating'' levels to
the total condensation energy is confirmed by a detailed
numerical analysis, see Sec.~\ref{secRichardson}.

Second, by numerically analyzing Richardson's equations
(\cite{Ric63,RS64,Ric65,Ric66}, see also a review in \cite{DR00}), we
find that in the regime $\lambda > \lambda^\ast$, the condensation
energy can be written as

\begin{equation}
  \label{eq:fullEcond}
  E_{cond} (\lambda) = E_{cond}^{BCS} (\lambda) + \Delta +
 \alpha (\lambda) E_{cond}^{pert} (\lambda) \; , 
\end{equation}
where $\alpha (\lambda)$ is a function of $\lambda$ of order unity.
(A rather similar, but not identical, form was obtained in Eq.(44) of
\cite{DS00} from a fit to numerical results for $E_{cond} (\lambda)$
obtained with the density matrix renormalization group.)  As will be
discussed in more detail in Sec.~\ref{secRichardson}, the first two
terms in Eq.~(\ref{eq:fullEcond}) represent the contributions of those
levels lying within $\Delta$ from $E_F$ (to be called ``condensed
levels''), while the last term is due to the remaining levels within
$\omega_D$ from $E_F$ (to be called ``fluctuating levels'').
According to Eq.~(\ref{eq:fullEcond}), the size-independent correction
to the BCS result (i.e. the leading order $1/N$ correction relative to
the extensive, bulk result) is at least $\Delta$.

The numerical analysis carried out in Sec.~\ref{secRichardson} and
App.~\ref{appseries} also give evidence that $E_{int}$ (and also
$E_{cond}$) is an analytical function on the positive real axis of
$\lambda$ with a radius of convergence around $\lambda=0$ of
approximately $1/\ln{N}$.  This is in agreement with our analytical
treatment of the perturbation series in
App.~\ref{appfactorial},\ref{appseries}.

The results for the condensation energy of grains with an odd number
of electrons are similar. In the ground state of an odd grain the
state at the Fermi level is occupied by a single electron. Due to the
considerations above, one does all the calculation neglecting this
level, and therefore, when the ground state energy is concerned, a
grain with an odd number of electrons is equivalent to a grain with an
even number of electrons with a noninteracting level spectrum which
does not contain the single level at the Fermi energy, and otherwise
identical. This change introduces only small quantitative changes in
the results above.

One way in which one can, in principle, measure the interaction  
energy of an ultrasmall superconducting grain is by a measurement of the 
specific heat. 
The interaction energy is then given by 
\begin{equation}
E_{int} =  \int_0^{\infty} [c_s(T) - c_n(T)] dT .
\label{specificcond}
\end{equation}
$c_{n(s)} = d\bar{E}_{n(s)}/dT$, where $\bar{E}_{n(s)}$ is the thermal 
average of the energy of a normal (superconducting) grain. 
While in macroscopic samples one obtains the leading order (extensive) term 
of the interaction 
energy by performing the above integral from zero to $T_c$, in ultrasmall 
grains, since the fluctuations involve states in the whole range of 
$E_F - \omega_D < \epsilon < E_F + \omega_D$, 
one has to replace the upper limit of the integral by 
$T_{max} \approx \omega_D$ in order to have a good estimate of $E_{int}$. 
At $T > T_{max}$, one expects that the interaction term 
in the Hamiltonian would play a negligible role, and ($\bar{E_s}$) 
and ($\bar{E_n}$) would be roughly the same. 

Another way to measure the interaction energy is by spin magnetization 
measurements, as we discuss in the next section.

\section{Spin magnetization of an ultrasmall grain}
\label{secmagnetization}

Since the condensation energy of an ultrasmall grain has contributions 
from all the levels 
within the range of $\omega_D$, in order to measure it one has to 
probe all the levels within this range. One way 
to do this is to put an ultrasmall, preferably
pancake-shaped grain in a magnetic field parallel
to the flat direction. One can then neglect orbital 
magnetization, and consider only the Pauli 
paramagnetism \cite{Pancake}. 

The interaction energy can then be obtained by: 

\begin{equation}
E_{int} = \int_0^{\frac{\omega_D}{\mu_B}} (M_n - M_s) dH ,
\label{magcond}
\end{equation}
where $M_{n(s)}$ is the magnetization of the normal (superconducting) grain.
This is a general thermodynamic identity, 
relying only on the fact that 
the electrons further than $\omega_D$ from $E_F$ are noninteracting, 
so that $M_n(H) = M_s(H)$ for $\mu_B H > \omega_D$. 
We now derive this relation for ultrasmall superconducting grains, and 
calculate the magnetization of such grains for $H \gg d/\mu_B$. 

We introduce the Zeeman term to the Hamiltonian (\ref{Hamiltonian}) 
changing $\epsilon_j \rightarrow \epsilon_j - \sigma \mu_B H$ (taking the 
$g$ factor to equal 2). Each eigenstate of the Hamiltonian 
(\ref{Hamiltonian}) 
is also an eigenstate of the modified Hamiltonian, with an energy  
$E_H = E_{H=0} - \mu_B H (n_\uparrow - n_\downarrow)$, where 
$n_\uparrow  (n_\downarrow) $ 
is the number of levels singly occupied by an electron with a spin in 
(opposite to) the direction of the magnetic field.

\begin{figure} 
        \begin{center}  
        \begin{picture}(1,0.618)
        \put(-0.0,-0.402){\psfig{figure=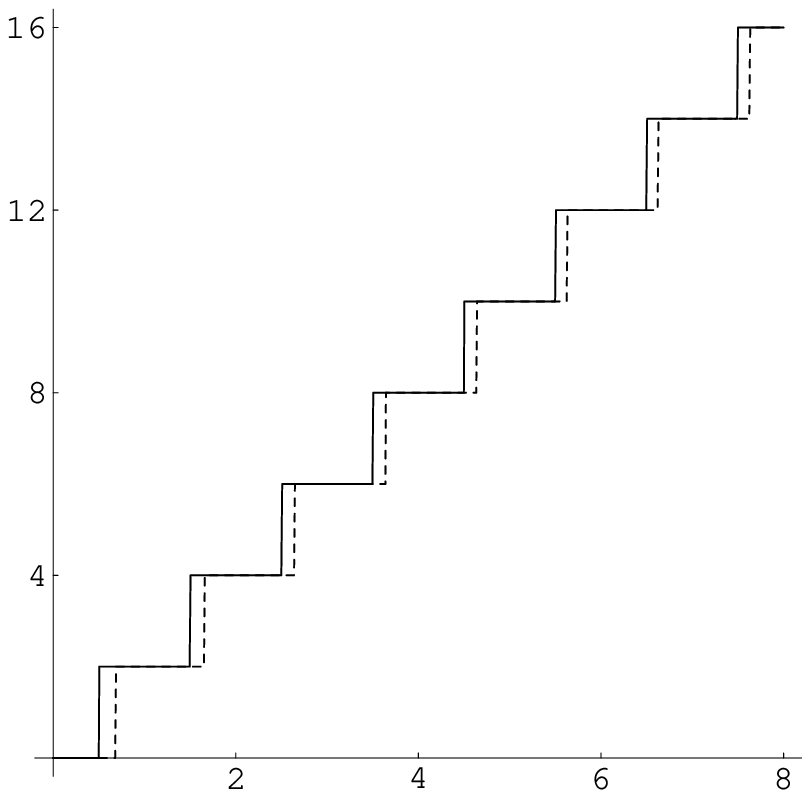,width=3in}}
        \put(0.05, 0.51){\makebox(0,0)[r]{{$M/\mu_B$}}}       
        \put(0.9,-0.45){\makebox(0,0)[lb]{$\mu_B$H/d}}
        \end{picture} 
        \end{center}
        \vspace{3.3 cm}
        \caption{Magnetization curve of normal (solid) and 
        superconducting (dashed) grains with equally spaced noninteracting 
        spectrum. The width of the 
        rectangles between the curves decreases with increasing magnetic 
        field due to the decrease of the second order term. The sum of all 
        the areas of the rectangles equals $E_{cond}/d$.}
\label{figmagsteps}
\end{figure}

We consider, as above, an ultrasmall grain with an even number of
electrons and neglect orders of $\lambda$ higher than two in the
calculations of the eigenstate energies below.  At $T=0$ and zero
magnetic field the ground state of the system has no broken pairs,
meaning there are no bare levels occupied with a single electron.
Of all the states with $l$ broken pairs, the one with the lowest
energy will be denoted $\psi_l$, and its energy $E_l$. One can show
that $\psi_l$ has all the $l$ levels closest to $E_F$ from above and
all the $l$ levels closest to $E_F$ from below singly occupied, while
all the other levels are not singly occupied. For $H \neq 0$ all the
electrons in the singly occupied levels will have their spin in the
direction of the magnetic field. In this case $E_l(H) = E_l(0) - 2 l
\mu_B H$.  For $T=0$ and finite $H$ the ground state of the system is
that $\psi_l$ with the smallest $E_l(H)$ of all $l$'s. While for a
large superconducting grain an abrupt transition from $l=0$ to $l =
\Delta/(\sqrt{2}d)$ occurs at $H=\Delta/(\sqrt{2}\mu_B)$
\cite{Clo62}, in an ultrasmall grain the number of broken pairs
in the ground state increases by one at a time as $H$ is increased
\cite{BD99}.
The magnetic field for which the transition of the ground state from
$\psi_{l-1}$ to $\psi_l$ occurs is denoted $H_l$.  For $H_l < H <
H_{l+1}$, $\psi_l$ is the ground state of the grain with ground state
energy $E_l(0) - 2 l \mu_B H$, and therefore the magnetization equals
$2 l \mu_B$. The magnetization is a step function in $H$, with equal
steps of magnitude $2 \mu_B$. One needs only to find the values of
$H_l$ to get the magnetization curve.  The above picture is true also
for a normal grain.  (By normal and superconducting grains we mean
here similar grains, with the same single particle noninteracting
spectrum, that differ only by the value of $\lambda$, which is zero
for the normal grain and finite for the superconducting grain.  The
relation of the above to a realistic situation is discussed below).
From its definition as the solution of
\begin{equation}
          E^{s/n}_l (H^{s/n}_l) = E^{s/n}_{l-1} (H^{s/n}_l) \; , 
\label{Hsnl}
\end{equation}
(for both pair-correlated or normal grains), $H_l^{s/n}$
is given by 
\begin{equation}
2 \mu_B H^{s/n}_l = E^{s/n}_l(0) - E^{s/n}_{l-1} (0) \; .
\end{equation}
It follows that
\begin{equation}
  \label{eq:suml}
  \sum_{l=1}^{l_{max}} 2 \mu_B H_l^{s/n} = E_{l_{max}}^{s/n} (0) - 
E^{s/n}_0  (0)
\; .
\end{equation}
Taking $l_{max} = \omega_D/d$
and subtracting the equation for normal grains from that for pair-correlated
ones, we find
\begin{equation}
  \label{eq:sumonl}
  \sum_{l=1}^{l_{max}} 2 \mu_B (H_l^{s} - H_l^{n} )
=  E^{n}_0  (0) - E^{s}_0  (0) = E_{int}
\; ,
\end{equation}
where we took $E_{l_{max}}^{s} (0) =
E_{l_{max}}^{n} (0)$, since at energies beyond
$l_{max} d = \omega_D$ the pairing interaction
is no longer operative. But, as can be seen
from Fig.~\ref{figmagsteps} (drawn for equally spaced normal and 
pair correlated grains), the sum on the left-hand side
equals the area between the solid and dashed lines, 
and hence also equals
the integral in Eq.~(\ref{magcond}).

Finding $H_l$ amounts to solving Eq.~(\ref{Hsnl}). 
We first assume that the noninteracting energy levels in the 
grain are equally spaced.
For a normal grain this equation then 
reduces to: $(2 l - 1) d - 2 \mu_B H = 0$, 
where the first term is the extra kinetic energy of the $l$ state compared 
to the $l-1$ state, and the second term is its gain in Zeeman energy.
In an ultrasmall superconducting grain one has to add the energy 
contributions due to $\hat{H_I}$ 
to the different ground states. 
To second order in $\lambda$, one can show, by using either Richardson's 
equations or perturbation theory, that the 
difference in the interaction energies of $\psi_l$ and $\psi_{l-1}$ is 

\begin{equation}
\lambda d+ \frac{1}{2}\left(\sum_{j=2l-1}^{N+l-1} 1/j+\sum_{j=2l}^{N+l} 
1/j\right) \lambda^2d\simeq \lambda d+\hat{\ln}(2l,N) \lambda^2d, 
\label{lenergy}
\end{equation}
where we define $\hat{\ln}(i,j) \equiv \sum_{k=i}^j 1/k$.
Therefore, the equation for $H_l$ is 

\begin{equation}
(2 l - 1) d + \lambda d + \hat{\ln}(2l,N) \lambda^2 d - 2 \mu_B H = 0 \; .
\label{Hl}
\end{equation}
The above equation is true for all $l<\omega_D/d$, while for larger $l$ 
the interaction term vanishes, and one obtains the same equation as for 
the normal grain. 

The first term in the equation reflects the kinetic energy cost of breaking 
the $l$'th pair, and is similar to the normal grain case. The second term 
reflects the 'direct' (Hartree) 
energy cost of breaking a pair, coming from the 
diagonal part of the interaction term in the 
Hamiltonian (\ref{Hamiltonian}). 
This term is 
not $l$ dependent, and therefore is not reflected in the susceptibility, as 
we shall see in the next section. 
The third term is the result of the two levels, the one $l$ below $E_F$, 
and the one $l$ above $E_F$, 
becoming blocked to pairing fluctuations. Its magnitude is a decreasing 
function of $l$, since as the levels are further from $E_F$ their 
contribution to the pairing fluctuations is smaller. This dependence on $l$ 
is reflected in the susceptibility. 

In Fig. \ref{figmagsteps} we plot the magnetization curve for a 
normal grain ($\lambda=0$) and a superconducting grain with the same 
equally spaced noninteracting spectrum.

Using Eq.~(\ref{eq:sumonl}), Eq.~(\ref{Hl}) 
and the expressions above it, one immediately
confirms the equality (to second order in $\lambda$) 
of the interaction energy, as
was calculated in App.~\ref{appfactorial}, 
and the integral in Eq.~(\ref{magcond}).

So far we considered the idealized case of grains with 
equally spaced energy levels. 
In order to relate to experiment, we now relax this assumption. 
It is not possible to 
see the effects of superconducting correlations 
on the condensation energy
by measuring only a single 
ultrasmall grain, since the fluctuations of the noninteracting 
energy levels cause 
larger shifts in the position of the $H_l$'s than those induced by the 
superconducting interaction. 
We therefore consider an ensemble of grains with the same 
noninteracting mean level spacing, $\tilde{d}$, 
and an energy spectrum that obeys GOE statistics. We assume that the 
pairing interaction constant in all the grains is the same, given by 
$\lambda \tilde{d}$ and calculate the mean spin magnetization of 
such an ensemble for $H \gg \tilde{d}/\mu_B$. 
For each grain, the equation (\ref{Hl}) for $H_l$ now becomes: 

\begin{equation}
\zeta_l + \lambda \tilde{d} + \hat{\ln}(2l,N) \lambda^2 \tilde{d} - 
2 \mu_B H_l = 0 \; .
\label{El}
\end{equation} 
$\zeta_l$ is the energy difference between the $l$'th level above the 
Fermi energy and the $l$'th level below it { in that grain}. 
The Hartree term is not affected 
by level statistics, and we neglect the 
change incurred by the second order term
due to the effects of level statistics,  since this change 
is small compared to its mean value. 
We approximate the second order term in Eq.~(\ref{El}) by 
$\lambda^2 \tilde{d} \ln[\omega_D/(2 \mu_B H)]$ 
(replacing $l$ inside the logarithm 
by its mean value, and replacing $\hat{\ln}$ by $\ln$). 
We then obtain, for a given magnetic field, for each grain, an equation for 
$l$, the number of broken pairs. It is given by the maximum $k$ that 
satisfies the equation: 
\begin{equation} 
2 \mu_B H - \lambda \tilde{d} - \lambda^2 \tilde{d}\ln[\omega_D/(2 \mu_B H)] 
\geq \zeta_k \; .
\label{maxl}
\end{equation}
The mean value of the magnetization of a grain at a given $H$ is therefore: 
\begin{equation}
\bar M_s (H)  = 
2 \mu_B^2 H/\tilde{d} - \lambda \mu_B - \lambda^2 \mu_B 
\ln[\omega_D/(2 \mu_B H)] \; .
\label{grainmag}
\end{equation}
The variation around the mean value is given by the variation of the
number of levels within the energy range given by the left side of
Eq.~(\ref{maxl}).  Since the level statistics of the grains is given
by GOE statistics, the variation in the magnetization of one grain is
approximately $\delta \bar M_s (H) = \mu_B \ln[2 \mu_B
H/\tilde{d}]/\pi^2$ (see e.g. \cite{Mel95}).  This variation is indeed
larger than the shift of the mean magnetization compared to that of a
normal grain [Eq.~(\ref{grainmag})], but in an ensemble of $n$ grains
the variation reduces as $1/\sqrt{n}$, while the shift in the mean
value does not change.

One can therefore, in principle, measure the interaction energy of
ultrasmall superconducting grains by measuring the magnetization of an
ensemble of such grains, and calculating the integral in
Eq.~(\ref{magcond}). While $M_s$ is measured, $M_n$ is given by the 
straight line starting from the origin with a slope equal to the 
measured ensemble magnetization at $H > \omega_D/\mu_B$.
The Hartree term 
in Eq.~(\ref{grainmag}) shifts the magnetization
of a pair-correlated grain 
relative to that of a normal grain 
by a constant, resulting in 
a parallel line not intersecting the origin. Its 
contribution to the integral is trivially $\lambda \omega_D$. The second 
order term changes the slope of the magnetization, and introduces a 
nonlinear correction to the normal Pauli susceptibility, which we 
discuss in the next section.  

The consideration of grains with an odd number of electrons would lead 
to similar results in the regime $\mu_B H \gg d$. 
The magnetization graph for an odd grain would be 
similar to that in Fig. \ref{figmagsteps}, only shifted by one unit down 
and half a unit to the left, not affecting the average quantities discussed.
 
\section{Re-entrance of the susceptibility} 
\label{secsusceptibility}

Measuring the interaction energy by a magnetization measurement might 
be a difficult task, since it requires very high magnetic fields, of the 
order of $\omega_D/\mu_B$. As an alternative, we propose here a 
susceptibility measurement which would reveal the presence of 
superconducting correlations in ultrasmall grains, and 
only requires magnetic 
fields of order $\tilde{d}/\mu_B$. 

Let $\chi^{s/n} (H,T) = \partial \bar M^{s/n} (H,T) / \partial H$
denote the spin susceptibility as function of magnetic field and
temperature, for a superconducting or normal grain, respectively.  Di
Lorenzo {\it et. al.} \cite{LFH+00} calculated $\chi^s(0,T)$, finding
that even for ultrasmall grains it has a minimum at $T \approx
\tilde{d}$, implying a re-entrant behavior as function of decreasing
$T$. Since this re-entrance differs from the monotonic increase
expected for the Pauli susceptibility $\chi^n(0,T)$ of normal grains,
they suggested that it could be a sensitive probe to detect
superconducting correlations in such grains.

\begin{figure}
        \begin{center}  
        \begin{picture}(1,0.618)
        \put(-0.028,-0.132){\psfig{figure=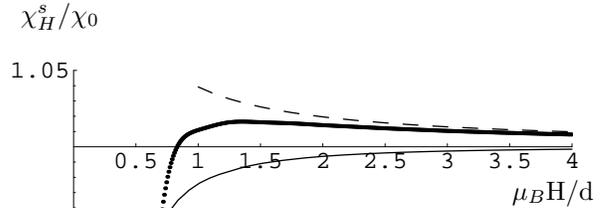,width=3in}}
        \put(0.13, 0.52){\makebox(0,0)[r]{{$\chi_H^s/\chi_0$}}}
        \put(0.85,0.2){\makebox(0,0)[lb]{$\mu_B$H/d}}
        \end{picture} 
        \end{center}
        \vspace{1.5 cm}
        \caption{Spin susceptibility as function of magnetic field 
        at $T=0$, for $\lambda = 0.28$, is shown. As $H$ decreases, 
	$\chi^s$ increases, until reaching a maximum of $1.02 \chi_0$ 
	for $H \approx 1.3 \tilde{d}/\mu_B$, implying a re-entrant 
	behavior. 
	$\chi^n(H)$ (thin solid line) 
	and the high field approximation obtained in 
	Eq.~(\protect{\ref{susceptibility}}) 
	(dashed line) are given for comparison. }
\label{figsusceptibility}
\end{figure}

In this section we discuss an analogous but  complementary
quantity, namely $\chi^s(H,0)$. 
We find that $\chi^s (H,0)$ has 
a maximum at $H \approx \tilde{d}/\mu_B$, 
and decreases 
as $1/H$ for $H \gg \tilde{d}/\mu_B$ (see Fig.~\ref{figsusceptibility}). 
Thus, $\chi^s(H,0)$ 
shows a re-entrant behavior in ultrasmall superconducting grains,
just as $\chi^s(0,T)$ does. Since this  again contrasts 
with the Pauli susceptibility $\chi^n(H,0)$ of normal grains
\cite{GE65,DMS73},  
measuring $\chi^s (H,0) $ as a function of 
$H$ could possibly serve as a sensitive probe to detect superconducting 
correlations in ultrasmall grains. 

For $H \gg \tilde{d}/\mu_B$ we use Eq.~(\ref{grainmag}) and obtain, 
to first order in $d/\mu_B H$: 

\begin{equation}
\chi^s = \chi_0 + \lambda^2 \mu_B/H .
\label{susceptibility}
\end{equation}
where $\chi_0 = 2 \mu_B^2/\tilde{d}$.
The susceptibility is a decreasing function of $H$, and 
the positive $1/H$-correction 
to the normal grain susceptibility, $\chi^n$ 
(for $H \gg \tilde{d}/\mu_B$, to first order in $d/\mu_B H$, 
one obtains $\chi^n = \chi_0$  \cite{GE65,DMS73}) 
is smaller than the 
leading, normal, term by $\lambda^2 \tilde{d}/2 \mu_B H = \lambda^2/2l$. 

The intuitive reason for why the correction is positive is as follows:
For a given magnetic field,
the magnetization of a pair-correlated grain is,
on the average, 
smaller than that of a normal grain (see Fig.~\ref{figmagsteps}), because
breaking pairs to increase the magnetization 
costs pairing energy. However, since the pairing
energy per extra pair decreases the further the 
pairs involved lie from $E_F$, the difference
between the two magnetization curves decreases
with increasing $H$. Consequently, it 
 requires a smaller $H$-increment to break the next pair
for a pair-correlated than a normal grain, implying
a larger susceptibility for the former. 

The result in Eq.~(\ref{susceptibility}) 
is already sufficient to establish 
the re-entrant behavior of the susceptibility $\chi^s(H,0)$, 
since as $H$ is lowered below $\tilde{d}$ and approaches  
zero, $\chi^s (H,0)$ decreases and approaches zero too, 
due to level repulsion. Precisely at $H = 0$ the susceptibility 
$\chi^s (H,0)$ of an odd grain has an additional 
$\delta (H)$-like peak due to the 
contribution of the single, unpaired electron at $E_F$; 
in fact, for finite $T$ it is the contribution of this 
unpaired electron that is responsible for the re-entrance 
of $\chi^s(0,T)$ predicted in Ref.~\cite{LFH+00}. However, 
for any non-zero $H$ the spin of this electron is fully 
aligned with the magnetic field and hence makes no 
contribution to $\chi^s(H>0,0)$. 

We now proceed with a calculation of $\chi^s$, which gives a quantitative 
estimate of the magnitude of the re-entrance effect. 
We consider an ensemble of odd and even grains.  
For a normal grain, $\chi^n(H)$  is proportional to the probability to have 
a pair of states ($l$ pair), $l$ above and $l$ below $E_F$ 
(denoted $-l$) 
separated by $\zeta_l \equiv \epsilon_l - \epsilon_{-l} = 2 \mu_B H$, 
and is given by \cite{GE65,DMS73}

\begin{equation}
\chi^n = \frac{2 \mu_B^2}{\tilde{d}} R(\frac{2 \mu_B H}{\tilde{d}})
\label{normalchi}
\end{equation} 
where $R(x)$ is the probability of finding two levels a spacing $x$ 
apart regardless of the position of the other levels \cite{Dys62} 
($R(x) = 1 - {\cal O}(1/x^2)$ for $x \gg 1$).
We now consider a single superconducting grain. The $l$-th pair of this 
grain would contribute, without pairing interaction, to the value 
of the spin susceptibility at $H = \zeta_l/(2 \mu_B)$. However, due 
to the extra energy cost of breaking a pair, which was discussed in the 
previous section, the $l$ pair contributes to the susceptibility at 
a higher magnetic field. This shift is specific to each grain, as it is 
a function of the energies of the noninteracting levels further than 
$i$ from the Fermi energy
(the levels closer than $l$ to $E_F$ are singly occupied and 
therefore do not contribute to the interaction energy).
While the energy of the $l$ pair is arbitrary, and later is taken to satisfy 
GOE statistics, we now 
make the approximation that the levels further than $l$ 
from $E_F$ are equally spaced with level spacing $\tilde{d}$, 
the first ones being $\tilde{d}$ 
apart from the $l$ levels. Due to the smallness of the 
fluctuations in the GOE 
ensemble we believe that the above approximation is not only proper for 
$H \gg \tilde{d}/\mu_B$, but gives fairly good results also for 
$H \approx \tilde{d}/\mu_B$. 
Under our approximation, which introduces 
a modification of Eq.~(\ref{lenergy}) due to the 
arbitrariness of the energies of the $l$ pair, we find that 
the $l$ pair will contribute to $\chi^s(H)$ at 
a $\zeta_l$-dependent field $H(\zeta_l)$ given by
\begin{equation} 
H = \frac{1}{2 \mu_B} \left[ \zeta_l + \lambda \tilde{d} + 
1/2 \cdot \left(\frac{1}{\zeta_l} + 
2 \sum_{j=1}^{N-l} \frac{1}{\zeta_l+j}\right) \lambda^2 \tilde{d}\right] .
\label{Hipair}
\end{equation}

\noindent 
Therefore, for an ensemble of ultrasmall superconducting grains 
as considered 

\begin{equation}
\chi^s(H) = (2 \mu_B^2/\tilde{d}) P(2 \mu_B H(\zeta)/\tilde{d})
\label{chis}
\end{equation} 
where $H(\zeta)$ is given by Eq.~(\ref{Hipair}) with $\zeta_l$ replaced by 
$\zeta$, and 
\begin{equation}
P(2 \mu_B H(\zeta)/\tilde{d}) = 
R(\zeta/\tilde{d}) (2 \mu_B dH/d\zeta)^{-1}. 
\end{equation}
The result for this calculation with $\lambda=0.28$ is given in 
Fig.~\ref{figsusceptibility}. 

For larger, but still small grains,  where $\Delta \geq \tilde{d}$, the 
spin susceptibility is very different at $H \lesssim \Delta/\mu_B$. 
However, similar calculations \cite{Sch00} to those leading to 
Eq.~(\ref{susceptibility}) show that for 
$\Delta^2/(\tilde{d} \mu_B) < H < \omega_D/\mu_B$ one obtains the same 
result as in Eq.~(\ref{susceptibility}).  
The reason essentially is that in this regime enough levels are 
singly occupied, so the energy levels involved in the interaction are 
sufficiently far for the perturbative treatment to be valid.

\section{Tunneling peak spectrum}
\label{sectunneling}

Superconducting correlations in ultrasmall grains are also 
reflected in their tunneling excitation spectrum.  

For a normal grain, the tunneling peak spectrum is simple, consisting 
of peaks at the single particle 
excitation energies of the grain. When the pairing interaction 
is present, the spectrum is much more complex, containing peaks at the 
energies of all the many body states of the grain with one  
electron added or removed. 
However, for a small coupling constant, most of these peaks are 
small (proportional to $\lambda^2$), and we do not consider them here. 
Instead, we consider only the ``primary'' peaks, i.e. 
those that survive as $\lambda \to 0$ and in this
limit correspond to the 
tunneling peaks of the normal grain.
A non-zero pairing interaction reduces their strength
by a factor of 
$1-\lambda^2$ and shifts their energy.
In this section we are 
concerned only with the energy shift of these primary 
peaks due to the 
pairing interaction. We show that this shift, being a decreasing function of 
energy, causes, for $\epsilon \gg d$, 
the mean spacing between the primary 
peaks to be smaller than in an 
equivalent normal grain, and therefore introduces a positive correction 
to their density. 

Consider a grain with equally spaced energy levels having an even
number ($2k$) of electrons and ground state $|\phi^G_{2k} \rangle$,
and consider the tunneling at positive energies, into any eigenstates
$|\phi_{2k+1} \rangle$ with $2k+1$ electrons.  We assume that the
Coulomb blockade energy is the same for the tunneling to all states
with $2k+1$ electrons, and henceforth neglect it, since we are
interested here only in energy {\em differences} between tunneling
peaks.

To first order in the tunneling Hamiltonian, tunneling will occur
whenever there is a finite matrix element between any state 
$|\psi_{2k+1}^s \rangle \equiv c^\dagger_s |\phi^G_{2k} \rangle$, 
where $s$ is an index labeling single noninteracting levels, and 
any eigenstate $|\phi_{2k+1}\rangle$ with $2k+1$ electrons.

If the grain is normal, $\lambda=0$, then the only relevant 
eigenstates with $2k+1$ electrons are those in  
which all levels up to $E_F$ are filled with two electrons, and one state, 
$s$, above $E_F$ is occupied by one electron.  
We define $| \phi_{2k+1}^s \rangle$ 
for either a superconducting or a normal grain 
as the lowest energy many body eigenstate of $2k+1$ electrons for which the 
state $s$ 
is singly occupied. For a normal grain  $| \psi_{2k+1}^s \rangle$ and 
$| \phi_{2k+1}^s \rangle$ are identical. 
The spectrum of tunneling peaks in a normal grain 
would therefore be identical to the 
noninteracting single-particle energy spectrum of the grain.
In a similar ultrasmall superconducting grain ($\lambda \neq 0$), pair 
fluctuations will affect the tunneling spectrum in three ways. 
(i) The primary peaks are shifted, and (ii) are 
reduced in magnitude due to the fact 
that the overlap of the states  $| \psi_{2k+1}^s \rangle$ and 
$| \phi_{2k+1}^s \rangle$ is smaller than one. 
(iii) Many small peaks emerge due to the small overlap 
(of order $\lambda$) between $| \psi_{2k+1}^s \rangle$ and all the 
other many body eigenstates with $2k+1$ electrons 
which are different from  $| \phi_{2k+1}^s \rangle$. 
We will not consider effects (ii) and (iii) here, and 
proceed with the calculation of the  
mean spacing between the primary peaks 
in ultrasmall superconducting grains, as a function of energy. 

The tunneling of an electron 
into the $l$'th level above the Fermi energy costs a 
total energy of
\begin{equation}
  E (\phi^l_{2k+1}) -   E (\phi^G_{2k}) = l d + \lambda d/2 + 
\lambda^2 d \hat{\ln}{(N,l)}/2 \; .
\label{tunnelingenergy}
\end{equation}
The first term is the kinetic energy contribution. The second term 
is the Hartree term, and the factor of $1/2$ is due 
to the fact that 
the number of pairs within $\omega_D$ below $E_F$ changes by, on average,  
$-1/2$ when an electron is added to the grain  
(because we have assumed that the band of
interacting electrons is spaced symmetrically about $E_F$, and $E_F$
shifts upward by half a unit of $d$ when an electron is added to the
grain).

Tunneling an electron into the $l$'th level also affects the
interaction energy, which to second order in $\lambda$ is reduced, due
to blocking level $l$, by $\lambda^2 d \hat{\ln}{(N,l)}/2 \simeq \lambda^2
\ln{(N/l)}/2$.  
(This second order term can be found, using e.g. perturbation theory, 
and calculating the difference in the second order interaction energy 
of a Fermi state with $2k$ electrons, and the same state up to a single 
electron added to the $l$'th level). 
Both the
kinetic energy and the Hartree term leave the distance between nearby
tunneling energies unchanged ($d$).  
However, the second order term becomes smaller with increasing energy,
and therefore the distance between tunneling energies is 
smaller than that of a similar normal grain. 
This reduction manifests itself in the mean spacing 
of the primary tunneling peaks in 
an ensemble of ultrasmall superconducting grains. One can obtain 
the mean primary 
peak spacing of such an ensemble by a similar procedure 
{ to the one} we took 
in Sec.~\ref{secmagnetization},~\ref{secsusceptibility} for 
the spin magnetization and susceptibility. 
Here we obtain the same result in a simpler way. 
We consider a grain with equally spaced energy levels as above. 
The difference between the 
tunneling energy to state $| \phi_{2k+1}^l \rangle$ and 
to state $| \phi_{2k+1}^{l+1} \rangle$ is approximately 
$d + \lambda^2 d [\ln{(N/(l+1))} - \ln{(N/l)}]/2 
\simeq d[1 - \lambda^2/(2l)]$. 
The mean density of primary peaks 
in an ensemble, for $l \gg 1, \epsilon \gg d$, including spin 
degeneracy, is therefore given by 

\begin{equation}
\bar{\cal N}(\epsilon) = \frac{2}{d} \cdot \frac{1}{1 - 
\lambda^2 d/(2\epsilon)} \; .
\label{dos}
\end{equation}

The functional behavior of the primary peak density resembles that 
of the magnetic susceptibility. In both cases the correction to the 
leading term reflects the change in interaction 
energy as levels further from the Fermi energy are blocked.

Similar considerations for negative energies (tunneling electrons out of 
the grain, going from the ground state of $2k$ electrons to states of 
$2k-1$ electrons) will result in a similar shift of the tunneling peaks, 
and the tunneling spectrum being symmetric around $E_F$. 

We now consider shortly the case of the tunneling process 
(at positive energies) changing a grain 
with an odd number of electrons to a grain with an even number of electrons. 
In the even to odd case described above, the change of the number of 
singly occupied noninteracting levels from zero to one induced an upward 
shift in the primary 
tunneling peak energies. Most primary tunneling peaks in the odd to 
even case correspond to a change in the number of singly occupied 
noninteracting levels from one to two, and therefore an upward 
shift in the tunneling peak energies 
similar to the even to odd case.
The only exception is the peak closest to the Fermi level, which is a 
result of tunneling into 
the singly occupied level, resulting 
in the even grain having no singly occupied levels. This induces a 
downward shift of this 
tunneling peak. The functional behavior 
of the mean peak spacing at $\epsilon \gg d$ will be similar to 
Eq.~(\ref{dos}) above.

We now relax the assumption of the noninteracting energy spectrum 
of the grain being equally spaced, and consider the regime 
$0 < \epsilon \ll d$. 
Due to the positive shift in energy of the tunneling peaks, as a result 
of the pairing interaction, $\bar{\cal N}(\epsilon)$ is smaller for ultrasmall 
superconducting grains compared to its value for similar normal grains 
(this can also be obtained by conservation of the number of primary peaks). 
This can be seen as a first sign for the onset of the gap in the density 
of states in a macroscopic sample. 

\section{Numerical analysis of Richardson's equations}
\label{secRichardson}

In this section we analyze the solutions of Richardson's equations 
(\ref{Richeq}) for the ground state of a grain with equally spaced 
energy levels. 

The above equations for the energy parameters can be studied numerically 
\cite{Ric66}. The energy parameters are in principle complex, and one 
can show that they are either real, or come in complex conjugate pairs.
If the ground state of the system is considered, it is found \cite{Ric66} 
that, for $\lambda \ll 1/\ln{N}$ all the energy parameters are real, 
monotonically decreasing functions of the coupling. 
Taking $N$ to be even, we  group all the 
energy parameters $E_\nu$ into 
pairs labeled by an index $a$, starting from
the largest two and counting downward:
$\{ E_{N+ 2 - 2a}, E_{N+1-2a} \}$, with $a = 1, \dots, N/2$.
(The case of odd $N$ can be
treated analogously, except that $E_1$ will remain unpaired then.) 
Each pair of energy parameters are real, until, for some $\lambda_a$ 
(Fig. 1 of Ref. \cite{Ric66}), the $a$'th pair 
becomes a pair of complex conjugate numbers, 
which they do in order of increasing $a$
(i.e.\ $\lambda_a < \lambda_{a+1}$).
At the transition point, $\lambda_a$, the two 
energy parameters are equal to the value of the lower energy parameter at 
$\lambda=0$: 

\begin{equation}
E_{N + 2 - 2a} = E_{N + 1 - 2a} = 2 \epsilon_{N + 1 - 2a} .
\label{firstpair}
\end{equation} 
Each energy parameter is an analytic function for 
$\lambda < \lambda_a$ and has a branch point at $\lambda_a$.

By solving Richardson's equations (\ref{Richeq}) for $\nu=N$ and $\nu=N-1$ 
with the conditions above (\ref{firstpair}) 
we find that 
\begin{equation}
\lambda_1 = 1/(\hat{\ln}{N} + a_1), 
\label{lambda1} 
\end{equation}
where 
$\hat{\ln}{N} = \sum_{j=1}^N 1/j$ and $0 < a_1 < 1$.

It would seem that the interaction energy, 

\begin{equation}
E_{int} = \sum_\nu (2 \epsilon_{\nu} - E_\nu),
\label{condeq}
\end{equation}
would be nonanalytical at this point. However, although 
$E_N, E_{N-1}$ have 
a branch point at $\lambda_1$, their sum is analytical at this point, 
due to the cancellation of the singularities. 
The analyticity of the sum, as well as the result in Eq.~(\ref{lambda1}), 
are derived \cite{Sch00} in the following way: 
We first define $\xi$ and $\eta$ to satisfy the 
equations: 
$E_{N} = 2 \epsilon_{N-1} + \xi + i \eta$ and 
$E_{N-1} = 2 \epsilon_{N-1} + \xi - i \eta$, using the fact that 
$E_{N}$ and $E_{N-1}$ are real (in which case $\eta$ is imaginary) or 
complex conjugates of each other
(in which case $\eta$ is real). We insert the above definitions 
in Richardson's equations for $\nu=N$ and $\nu=N-1$, and obtain 
equations for $\xi$ and $\eta^2$ as function of $\lambda$
(similar equations are given in \cite{Ric66,DR00}). 
We then  expand $\xi$ and $\eta^2$ in a series in 
$\delta \lambda = (\lambda - \lambda_1)$, 
and solve the equations for each order of $\delta \lambda$ separately. 
$\lambda_1$ is obtained by the solution to the zero'th order, 
and yields Eq.~(\ref{lambda1}). 
Solving for the next orders we find that the coefficients, for both 
$\xi$ and $\eta^2$, of $(\delta \lambda)^i$ are, up to a factor of 
order unity, $(\ln{N})^{2i}$. This results in 
$\xi$ and $\eta^2$ being in fact expanded as a series in 
$\delta \lambda \cdot (\ln{N})^2$. 
This result both reflects a scale of 
$1/(\ln{N})^2$ near $\lambda = 1/\ln{N}$, which we will return to later, 
and shows that the sum $E_N + E_{N+1}$ can indeed be expanded 
perturbatively in $\delta \lambda$. 
This result also suggests that $E_{int}(\lambda)$ 
is analytical on the positive real 
axis. The only points at which 
one can suspect nonanalyticities to occur are the different 
$\lambda_a$'s. The reasonable  assumption 
that for $a>1$ the behavior of $E_{int}$ near $\lambda_a$ 
is similar to that near $\lambda_1$ discussed above, 
i.e. that the sum of the 
singular energy parameters is analytical, leads to the analyticity 
of $E_{int}$ on the positive real axis. This is also supported by contour 
integration in the complex $\lambda$ plane, as we discuss in 
App~\ref{appseries}.

\begin{figure}
        \begin{center}  
        \begin{picture}(1,0.618)
        \put(0.02,-0.95){\psfig{figure=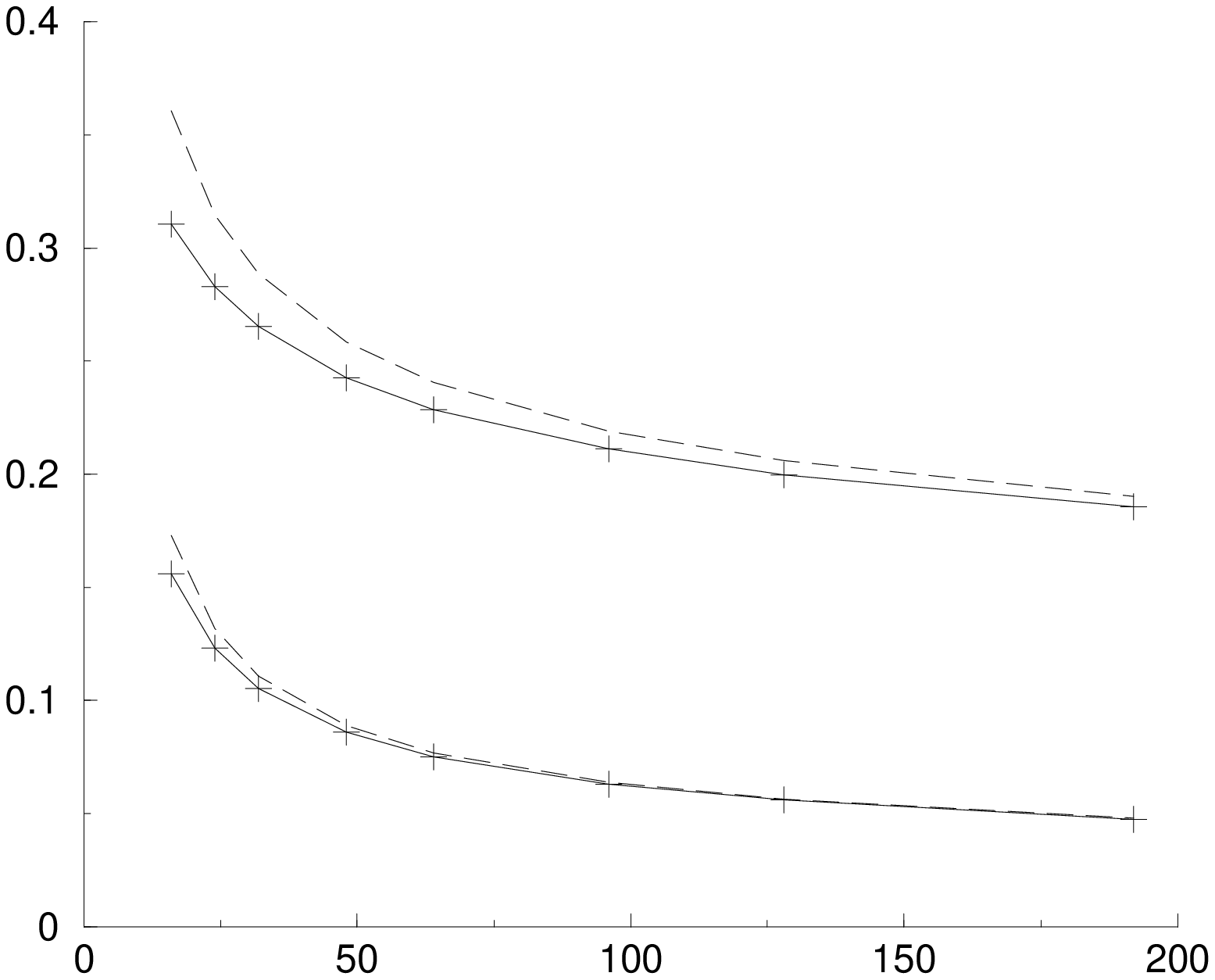,width=2.5in}}
        \put(0.0,-0.1){\makebox(0,0)[lb]{\psfig{figure=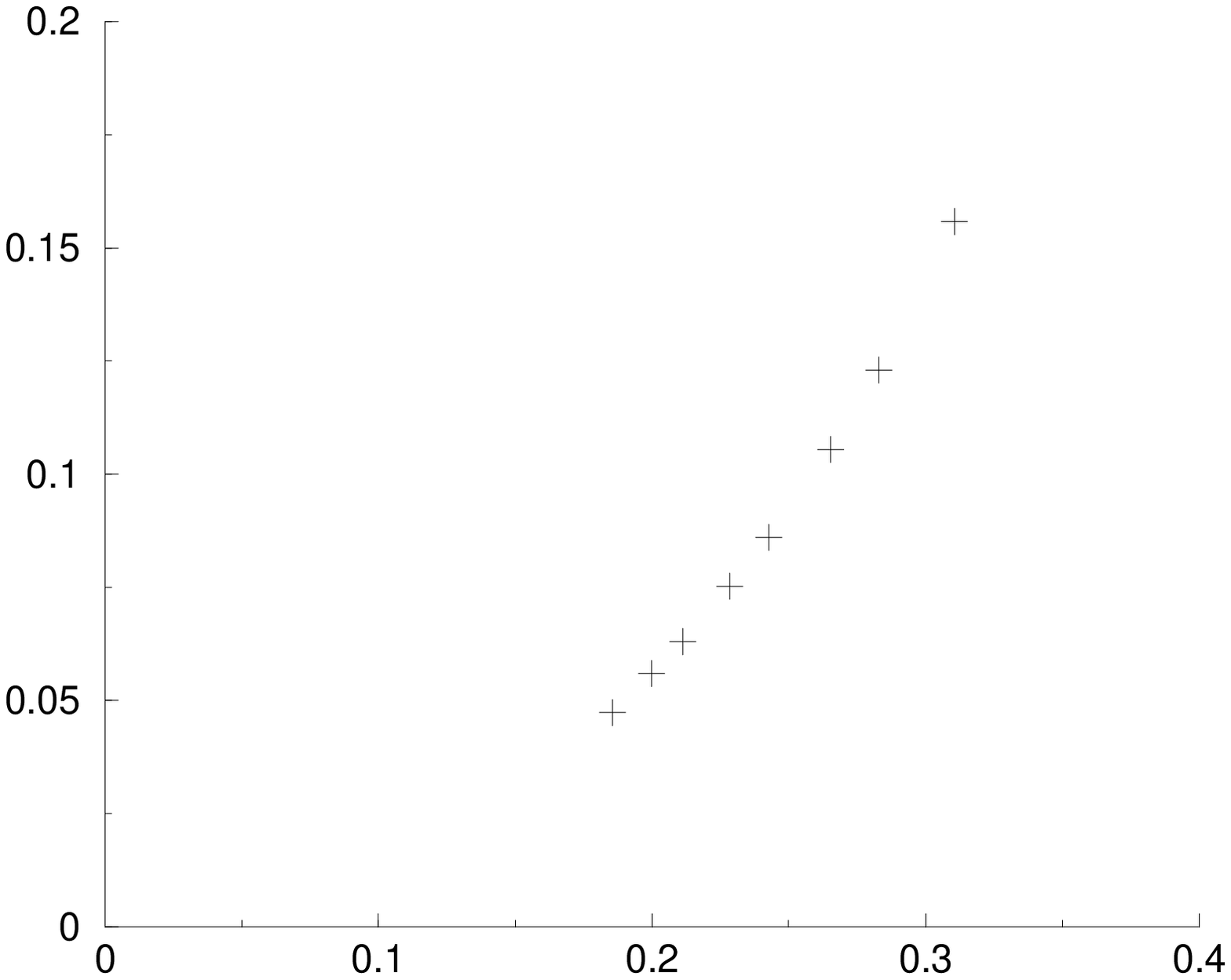,width=2.5in}}}
        \put(0.8, -0.97){\makebox(0,0)[t]{{$N$}}}
        \put(-0.01, -0.33){\makebox(0,0)[r]{{$\lambda$}}}
        \put(0.05, -0.4){\makebox(0,0)[r]{{(Re,Im)}}}
        \put(0.78, -0.12){\makebox(0,0)[t]{Re $\lambda$}}        
        \put(0.05, 0.48){\makebox(0,0)[r]{Im $\lambda$}}        
        \put(0.45, -0.17){\makebox(0,0)[t]{(a)}} 
        \put(0.45, -0.99){\makebox(0,0)[t]{(b)}}
        \end{picture} 
        \end{center}
        \vspace{8.0 cm}
        \caption{(a) The singularity point of the complex coupling 
        parameter which is closest to the origin is plotted 
        for (from top right to bottom left) $N=16,24,32,48,64,96,128,192$. 
        Not plotted is their mirror image across the x axis, with negative 
        imaginary parts.
        (b) The real (top) and imaginary (bottom) parts of the 
        complex coupling parameter at the above singularity points are 
        plotted as function of $N$. The dashed lines are fits, $1/\ln{N}$
        for the real part, and $1.33/(\ln{N})^2$ for the imaginary part. 
        The error in both graphs is 0.002 (not plotted).  }
\label{figsingular}  
\end{figure}

However, $E_{int}$ is not an analytical function in the whole complex 
$\lambda$ plane. 
By numerically analyzing Richardson's equations as a function of complex 
$\lambda$, we find that the interaction energy has singular points in 
the complex plane, the closest to the origin occurring at  
approximately 
\begin{equation}
\lambda_{sing}(N) = 1/\ln{N} \pm 1.33 i/(\ln{N})^2. 
\label{lambdas}
\end{equation}
In Fig. \ref{figsingular} we plot the 
real and imaginary parts of the closest singular points as function of 
$N$, with the corresponding fits.

This numerical result suggests that the radius of convergence of the 
perturbation series for the interaction energy around 
$\lambda = 0$ is roughly 
$1/\ln{N}$, in agreement with our analytical treatment of the perturbation 
series in App.~\ref{appfactorial},\ref{appseries}.
It also reflects a convergence radius of the order 
of $1/(\ln{N})^2$ for the sum $E_N + E_{N-1}$ around $\lambda = \lambda_1$, 
in agreement with the $1/(\ln{N})^2$ scale mentioned above.

The second part of this section is devoted to establishing a 
connection between the analytical properties of Richardson's energy 
parameters and the BCS theory. In particular, we show that 
in the regime where $d \ll \Delta$ the BCS 
result for the condensation energy (\ref{eq:EcondBCS}) is closely related 
to the singular contribution of the complex energy parameters to the 
condensation energy, 
and that the points, on the positive 
real axis of $\lambda$, at which the energy parameters become complex, 
are related to the values of $\lambda$ at which additional states become 
``condensed'' (i.e. come to lie within $\Delta$ of $E_F$). 

We separate the energy parameters to two groups, $R_\lambda$ and
$C_\lambda$, containing those energy parameters which, for a given
$\lambda$, are real or complex, respectively.  We define the
contribution of the complex energy parameters to the condensation
energy (disregarding the Hartree term): 

\begin{equation}
E_{cond}^{comp} = \sum_\nu^{C_\lambda} (2 \epsilon_\nu - \lambda - E_\nu)
\label{eqcontributions}
\end{equation}
and find 
that $|E_{cond}^{comp} - (\Delta^2/2 d + \Delta)| < 3 d \,\, $ for 
$ 8 < N < 1024$, for all $\lambda < 0.3$. 
The relative correction decreases with $N$ and $\lambda$, and is $0.1$ 
percent for $N=1024$ and $\lambda=0.3$. 

For those energy parameters which have already become complex at a
given $\lambda$, we can separate the contribution of $E_{2N+2-2a} +
E_{2N+1-2a}$ to the condensation 
energy into two parts, one containing the decrease which each
$E_{2N+2-2a} + E_{2N+1-2a}$ underwent as $\lambda$ is increased from 0
to $\lambda_a$ (the perturbative regime), the other containing 
its further decrease for
$\lambda > \lambda_a$ (the singular regime).

\begin{figure}
        \begin{center}  
        \begin{picture}(1,0.618)
        \put(-0.02,-0.12){\psfig{figure=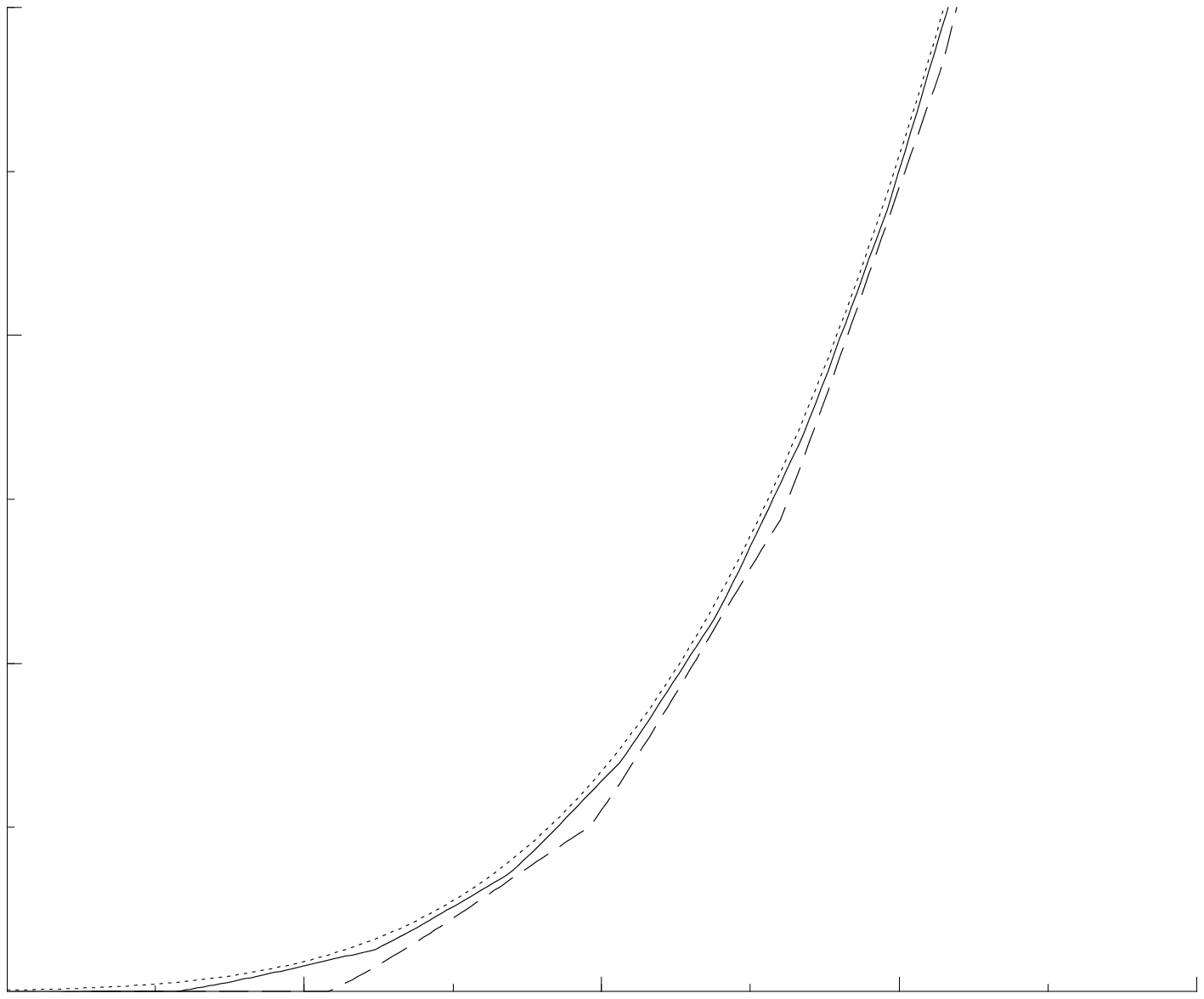,width=2.5in}}
        \put(0.01,-0.99){\makebox(0,0)[lb]{\psfig{figure=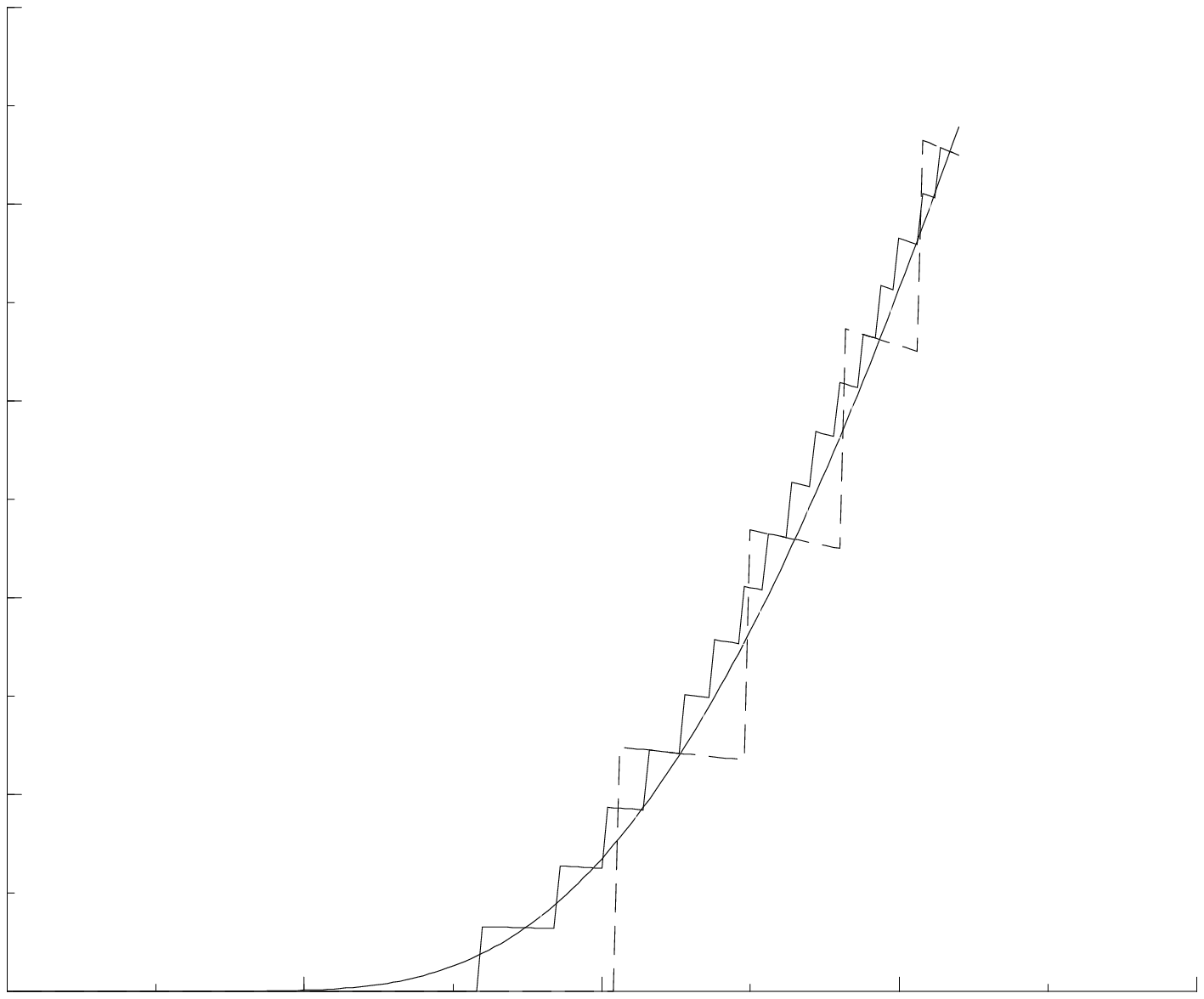,width=2.5in}}} 
	\put(0.44, -0.17){\makebox(0,0)[t]{(a)}}
        \put(0.46, -1.07){\makebox(0,0)[t]{(b)}}
        \put(0.75, -0.12){\makebox(0,0)[t]{{$\lambda$}}}
        \put(0.2, 0.57){\makebox(0,0)[r]{{$E_{cond}^{sing}/(2N^2d)$}}}
	\put(0.06, 0.3){\makebox(0,0)[r]{0.001}}       
        \put(0.25, -0.11){\makebox(0,0)[t]{0.2}} 
        \put(0.61, -0.11){\makebox(0,0)[t]{0.3}}
        \put(0.06, -0.71){\makebox(0,0)[r]{0.2}}       
        \put(0.06, -0.465){\makebox(0,0)[r]{0.4}}       
        \put(0.46, -0.98){\makebox(0,0)[t]{0.2}} 
        \put(0.83, -0.98){\makebox(0,0)[t]{0.4}}
        \put(0.78, -1.03){\makebox(0,0)[t]{$\lambda$}}
        \put(0.2, -0.27){\makebox(0,0)[r]{$E_{comp}^{pert}/(2Nd)$}}
        \end{picture} 
        \end{center}
        \vspace{8.5 cm}
        \caption{The different contributions of the complex energy 
	parameters to the condensation energy are plotted. (a) 
	We show that $E_{cond}^{sing}$ is well approximated 
	by the BCS result, by plotting $E_{cond}^{sing}/(2N^2d)$ 
	for $N=64$ (dashed line) and $N=128$ (solid line) 
	and the function $exp(-2/\lambda)$ (dotted line) for comparison. 
	Plots for larger $N$ are not drawn since they are 
	indistinguishable from $exp(-2/\lambda)$ in the resolution of 
	this figure. (b) $E_{comp}^{pert}/2Nd$ for $N=64$ (dashed 
	line, large steps) and $N=256$ (solid line, small steps) is 
	plotted and compared to $exp(-1/\lambda$) (smooth solid line), 
	showing that $E_{comp}^{pert} \simeq \Delta$. Steps occur 
	at points $\lambda_a$ in which pairs of energy parameters 
	become complex.}
\label{figBCS}
\end{figure}

Defining, for each energy parameter $E_\nu$, its value at the branch point 
as $E_{\nu}^b \; $ (by Eq.~(\ref{firstpair}) 
$E_{2l}^b = E_{2l-1}^b = 2\epsilon_{2l-1}$), we write 
$E^{comp}_{cond} = E^{sing}_{cond} + E^{pert}_{comp} $, where 
\begin{equation}
E_{cond}^{sing} =  \sum_\nu^{C_\lambda} (E_\nu^b - E_\nu) \; , \quad 
E_{comp}^{pert} = \sum_\nu^{C_\lambda} (2 \epsilon_\nu - \lambda - E_\nu^b) 
\; . 
\label{eqtwocontributions}
\end{equation}
$E^{sing}_{cond}$ is the singular contribution to $E_{cond}$ 
or $E_{int}$. (Note that the Hartree term is subtracted
in $E^{pert}_{comp}$ when calculating the
condensation energy; for the analogous calculation of the
interaction energy, this subtraction should be omitted.) 
Remarkably, our numerical analysis shows 
that this singular contribution 
is well approximated by the BCS expression for the 
condensation energy 
($E_{cond}^{sing} = \Delta^2/2d (1 + {\cal O}(d/\Delta)^2)$ for $64<N<1024$ 
for all $\lambda<0.3$) as is shown in Fig.~\ref{figBCS}(a).  
This suggests that 
the BCS approximation is equivalent to considering {\it only} 
the singular contribution to the 
condensation  
energy. By taking first the 
limit $N \rightarrow \infty$, one would indeed get the singular point of 
the interaction energy to be at the origin (\ref{lambdas}), 
and no contribution from the 
perturbative terms. In this limit the Hartree term in the reduced 
Hamiltonian (\ref{Hamiltonian}) vanishes, and the 
condensation energy equals the interaction energy. 

These results suggest that the $1/N$ correction to the BCS result in the 
large $N$ limit is at least $\Delta$. However, one also has to add 
the contribution of the energy parameters in group $R_\lambda$, 
which correspond to the 
levels between $E_F - \omega_D$ and $E_F - \Delta$. This contribution 
would give an additional correction of order $\lambda^2 \omega_D$. 
Summing the above three contributions to $E_{cond}$, we obtain 
Eq.~(\ref{eq:fullEcond}). 

Since $E^{comp}_{cond} \simeq \Delta^2/2d + \Delta + {\cal O}(d)$, 
and $E_{cond}^{sing} \simeq \Delta^2/2d + {\cal O}(d)$, 
we find that   
$E_{comp}^{pert} \simeq \Delta$ [see Fig.~\ref{figBCS}(b)]. 
This implies, since $2 \epsilon_\nu - E^b_\nu$ equals $2 d$ for even
$\nu$ and $0$ for odd $\nu$, the approximate equation 
\begin{equation}
2 n_C (\lambda) \simeq \frac{\Delta(\lambda)/d}{(1 - \lambda)}
\label{lambdai}
\end{equation}
where $\Delta (\lambda)$ is 
given in Eq.~(\ref{eq:bulkgap}), and 
$n_C (\lambda)$ is the number of pairs of energy parameters 
$\{E_{N+2-2a},E_{N+1-2a} \}$
that have already turned complex for the given $\lambda$. Remarkably,
Eq.~(\ref{lambdai}) tells us that the associated number of bare levels
$\epsilon_{N+2-2a}$ and $\epsilon_{N+1-2a} \;$, namely
$2n_C(\lambda)$, is just the number of bare levels within $d$ of   
$\Delta$ (up to a factor close to unity), i.e.\ the number of what we
have called ``condensed levels''.  The reason for this nomenclature 
now becomes apparent, since we have just established that the   
singular, BCS-part of $E_{cond}$ arises precisely from those energy parameters
that have evolved from these $2n_C(\lambda)$ bare levels.

\begin{figure}
        \begin{center}  
        \begin{picture}(1,0.618)
        \put(-0.08,-0.232){\psfig{figure=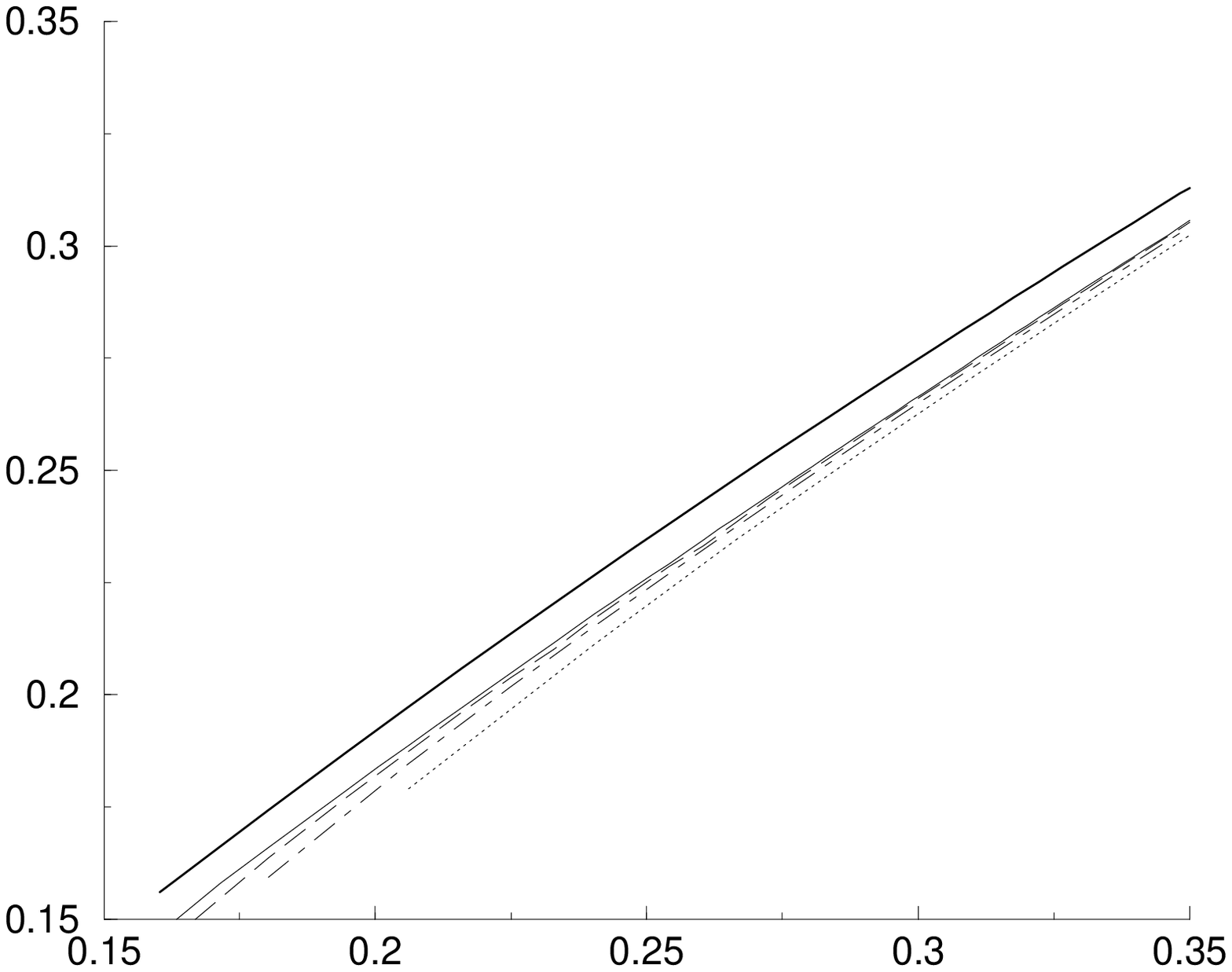,width=3.375in}}
        \put(0.13,0.25){\makebox(0,0)[lb]{\psfig{figure=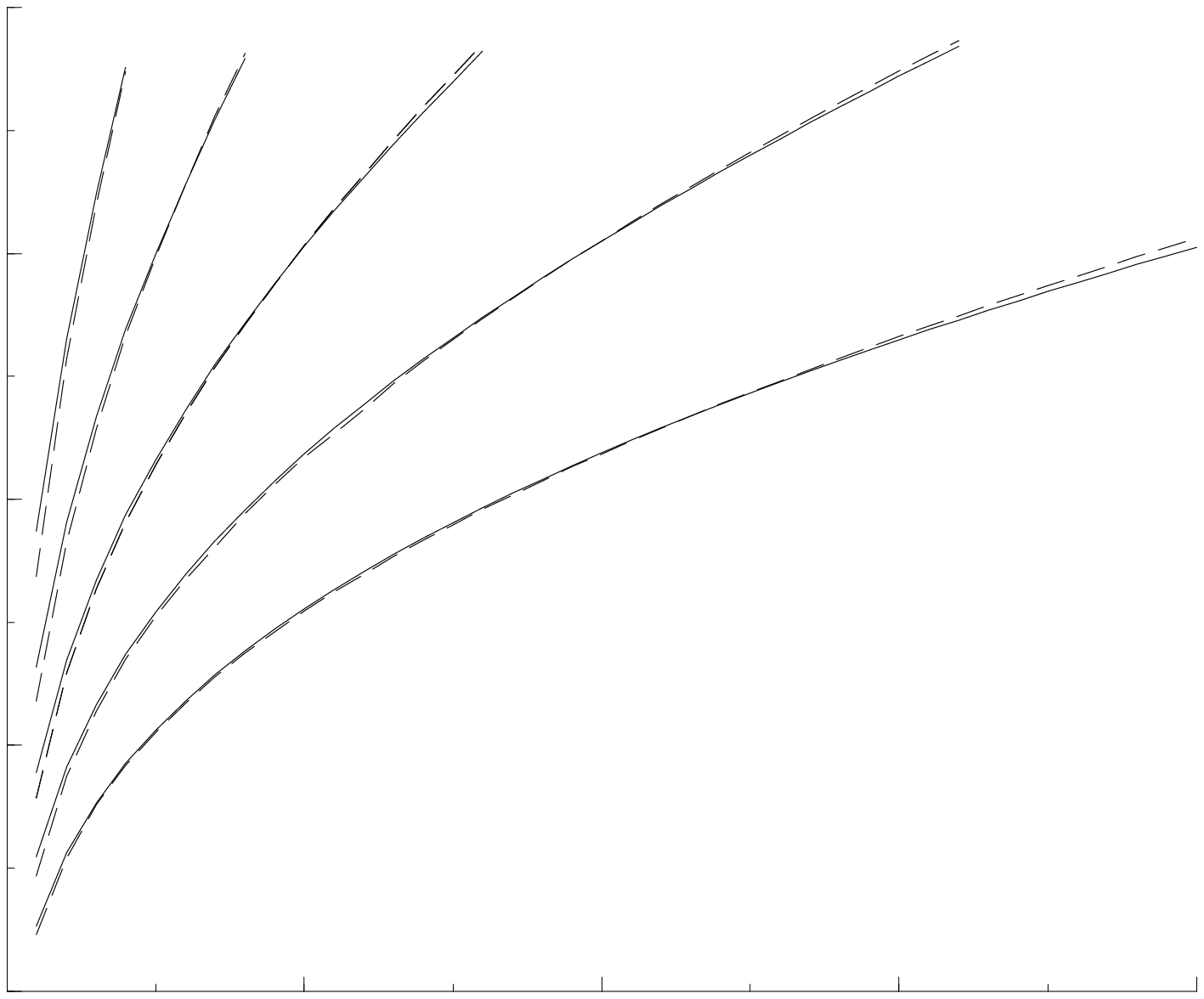,width=1.35in}}} 
        \put(-0.02, 0.58){\makebox(0,0)[r]{{$\lambda_a$}}}
        \put(0.12, 0.63){\makebox(0,0)[r]{{$\lambda_a$}}}
        \put(0.12, 0.35){\makebox(0,0)[r]{\small{0.17}}}       
        \put(0.12, 0.53){\makebox(0,0)[r]{\small{0.27}}}       
        \put(0.24, 0.24){\makebox(0,0)[t]{\small{10}}} 
        \put(0.355, 0.24){\makebox(0,0)[t]{\small{20}}}
        \put(0.47, 0.24){\makebox(0,0)[t]{\small{30}}}
        \put(0.54, 0.245){\makebox(0,0)[t]{k}}
        \put(0.85,-0.3){\makebox(0,0)[lb]{$1/\ln{\frac{N}{a}}$}}
        \end{picture} 
        \end{center}
        \vspace{2.5 cm}
        \caption{The bottom four curves are plots of $\lambda_a$ as a 
        function of $1/\ln{(N/a)}$ computed by changing $a$ for fixed (from 
        bottom up) $N=128,256,512,1024$. The topmost (thick, solid) 
	curve is the solution 
        to Eq. (\protect{\ref{lambdai}}).  In the inset we plot $\lambda_a$ 
        as a function of $a$ for (from top left to bottom right)
        $N=64,128,256,512,1024$. The numerical curves (dashed) are each fit 
        with the solution to 
        Eq. (\protect{\ref{lambdai}}) minus $0.055/\ln{N}$ (solid lines).
        The error in $\lambda_a$ is $0.0002$, and is not drawn.  }
        \vspace{0.8 cm}
\label{figlambdai}
\end{figure}

If we solve Eq.(\ref{lambdai}) for $\lambda$, the result
gives a function, say $\lambda_{approx}(n_C)$,
that actually depends on $N$ and $n_C$ only via the ratio $N/n_C$,
and that is plotted as function of $1/ \ln(N/n_C)$ in 
Fig.~\ref{figlambdai} (thick  
bold line).  This function can be used to approximately predict, for
given $N$ and $n_C$, at which value of the coupling constant the
$n_C$-th pair of energy parameters will become complex. For
comparison, we plot in Fig.~\ref{figlambdai} 
also the actual value at which this 
happens, say $\lambda(n_C)$, for several values of $N$ (non-bold
lines, obtained by numerically solving Richardson's equations).  We
find that the difference can be well fit by  

\begin{equation}
\lambda_{approx} (n_C) - \lambda (n_C) \simeq 0.055/ \ln N,
\label{eq:oneparameterfit}
\end{equation}
implying that the
approximate prediction becomes very accurate for $N \to \infty$.
Eq.~(\ref{eq:oneparameterfit}) represents a one-parameter fit that can 
be used for any combination of $N$ and $n_C$, and the quality of which
is illustrated in more detail in the inset of Fig.~\ref{figlambdai}.

Expanding $\lambda_{approx}(n_C)$ to first order
in $\ln (N/n_C)$, we find that
\begin{equation}
\lambda (n_C) =  1/\ln{(N/n_C)} \simeq \frac{1}{\ln{N}} + 
\frac{\ln{n_C}}{(\ln{N})^2} .
\label{kpar}
\end{equation}
This equation shows, once again, that the scale of
$1/(\ln N)^2$ is present also for $\lambda > \lambda^\ast$,
both in BCS theory and in the analytical properties
of Richardson's equations.

\section{Conclusions}
\label{conclustion}

Though many superconducting properties are limited to grains large enough 
such that $d<\Delta$, pair correlations exist also in ultrasmall grains, where 
$d>\Delta$. We calculated the effect of pair correlations on the 
condensation 
energy, spin magnetization and the tunneling spectrum of ultrasmall 
superconducting grains. We found that the contribution of pair fluctuations 
to the condensation energy is much larger than the BCS result even for 
grains in which $\Delta>d$, and that 
the condition for the validity of the BCS 
approximation for calculating the condensation energy is 
$\Delta>\sqrt{\omega_D d}$.
The interaction energy of ultrasmall grains can, in principle, be 
experimentally obtained through measuring their spin magnetization, which 
was calculated above. The pair correlations result in a 
positive correction to the differential 
spin susceptibility which is proportional to $\lambda^2 d/(\mu_B H)$
for $\mu_B H \gg d$, and a positive 
correction to the 
mean tunneling peak density which is proportional to 
$\lambda^2 d/\epsilon$ for $\epsilon \gg d$. 
The differential 
spin susceptibility at $T=0$ of ultrasmall superconducting grains shows 
a re-entrant behavior as a function of $H$, which 
could serve a sensitive probe for the existence of superconducting 
correlations in such grains. 
We argued that the interaction energy is an analytic function of the 
coupling parameter, with a convergence radius of approximately $1/\ln{N}$.
We showed that the BCS result for the { condensation} energy can be 
obtained from the singular part of Richardson's energy parameters, and that 
the correction to the BCS result in the regime where $d \ll \Delta$ is 
at least $\Delta$.

M.S. would like to give special thanks to Boris Laikhtman and Iddo
Ussishkin for enlightening discussions.  We would also like to thank 
B. L. Altshuler, 
F. Braun, A. M. Finkel'stein, U. Gavish, M. Kirson, Z. Ovadyahu, and 
D. Prober 
for useful discussions.  This work was supported by the Israel Academy
of Science and by the German-Israeli Foundation (GIF).

\appendix

\section{Accuracy of the energy approximation} 
\label{appaccuracy}

In this appendix we show that the relative accuracy of the approximations 
in Eqs.~(\ref{delEnuapprox}), (\ref{condmatveev}) is 
as stated in Eq.~(\ref{relativelnall}). 
In this appendix we take $d=1$.

From Eqs.~(\ref{anu0}) and (\ref{anu}),
$\delta a_\nu = a_\nu - a^0_\nu$  is given by: 

\begin{eqnarray} 
\delta a_\nu = - \sum_{j=1 (\neq \nu)}^{2N} 
\frac{\delta E_\nu}{(2 j - 2 \nu + \delta E_\nu) 
(2 j - 2 \nu) } + 
\nonumber \\
\sum_{\mu=1 (\neq \nu)}^N \frac{2 (\delta E_\mu - 
\delta E_\nu)}{(2 \mu - 2 \nu + \delta E_\nu - \delta E_\mu) (2 \mu - 2 \nu)} 
\; .
\label{deltaanu}
\end{eqnarray}

For $\lambda < 1/\ln{N} - c/(\ln{N})^2$, 
we assume (and later check for consistency) that 
$0 < \delta E_{\nu} < 1/c \;$ 
for all $\nu$. Separating each sum in Eq.~(\ref{deltaanu}) into 
a sum over the levels above and below $\nu$, 
one can see that $|\delta a_{\nu}| < (1/c) \cdot 3 \cdot 2 \sum_s 
[2 s (2 s -1)]^{(-1)} \Rightarrow |\delta a_{\nu}| < 6/c$ 
(actually a more careful 
treatment can reduce the numerical factor multiplying $1/c$ to be of order 
unity, but this is of no importance here).

Therefore, one can write: 

\begin{equation} 
\delta E_{\nu} = 
\frac{\lambda}{1 - \lambda \, a_{\nu}^0 + \lambda b_{\nu}(\lambda)/c} \; , 
\label{delEnuexacteq}
\end{equation} 
where $|b_{\nu}(\lambda)| < 6$. Manipulating 
the above equation, we obtain 
$\delta E_{\nu} = \delta E_{\nu}^0 + R_{\nu}$ where 

\begin{equation}
R_{\nu} = - \frac{\lambda^2 b_{\nu}(\lambda)}{c \cdot (1 - \lambda a_{\nu}^0 + 
\lambda b_{\nu}(\lambda)/c) \cdot (1 - \lambda a_{\nu}^0)} \; .
\label{Rnu}
\end{equation}
The relative accuracy is therefore: 

\begin{equation}
r_{\nu}(\lambda) \equiv |\frac{R_{\nu}(\lambda)}{\delta E_{\nu}(\lambda)}| = 
\frac{\lambda |b_{\nu}(\lambda)|}{c \cdot (1 - \lambda a_{\nu}^0)} \; .
\label{rnulam}
\end{equation}

Consider now $\lambda < 1/\hat{\ln}{N} - c/(\hat{\ln}{N})^2$, and $\nu=N$ 
(the highest energy parameter). Since $a_N^0 = (\hat{\ln}{N} + 
\hat{\ln}{N-1})/2 \simeq \hat{\ln}{N}$ we see that $r_N < 6/c^2$.

One can obtain from Eq.~(\ref{anu0}) that $a_\nu^0$ 
increases monotonically with  $\nu$, 
and by differentiating Eq.~(\ref{rnulam}) with respect to $a_{\nu}^0$, 
that $r_\nu$ 
increases monotonically with  $a_\nu^0$. 
We conclude that $r_{\nu} < 6/c^2$ for all $\nu$, 
and that therefore $ \delta E_{\nu} = 
\delta E_{\nu}^0[1 +{\cal O}(1/c^2)]$ for all $\nu$. 
In the same way one finds that $\delta E_{\nu} < 1/c$ for all $\nu$, 
consistent with our assumption above. 

By summing over all $\nu$, one shows that 

\begin{equation}
E_{int} = E_{int}^0 (1 + {\cal O}(1/c^2)) ,
\end{equation}
where $E_{int}^0$ is given in Eq.~(\ref{condmatveev}). 

To show that the accuracy of the above expression is $1/(\ln{N})^2$ in the 
regime $\lambda < 1/(2 \ln{N})$ one has to assume (and later show 
consistency) that 
$\delta E_{\nu} < 1/\ln{N}$ for all $\nu$, and proceed as above.

\section{Approximate formula for the $m$'th order term of the 
interaction energy}
\label{appfactorial}

In this appendix we expand Eq.~(\ref{condmatveev})
for the approximate interaction energy $E_{int}^0$ 
in powers of $\lambda$, 
and analyze the convergence properties of the resulting
series. We begin by showing that 
the order-$\lambda^m$ contribution to $E_{int}^0$
has the form 
\begin{eqnarray}
E_{int}^{0(m)} &  = & d \, \lambda^m \sum_{\nu=1}^N (a_\nu^0)^{m-1} 
\label{morder}
\\
& = & \left\{ 
\begin{array}{ll}
d \, \lambda^m \, b_m N (m-1)!  &  \quad \mbox{for} \quad m < \ln{N} \\
d \, \lambda^m \, 
b'_m (\ln{N})^{(m-1)}  &  \quad \mbox{for} \quad m > \ln{N}  \; ,
\end{array} \right.
\label{limitm}
\end{eqnarray}
where $b_m$ 
and $b'_m$ 
are constants of order unity,
and then analyze the consequences of this result.
Since we are not interested here in numerical factors of order unity, we will 
allow ourselves to make some crude approximations.

The expression obtained (\ref{condmatveev}) for the { interaction} 
energy can 
be written as: 

\begin{equation}
E_{int}^0 = \sum_m E_{int}^{0(m)} = d \sum_m [\lambda^m \sum_{\nu=1}^{N} 
(a_{\nu}^0)^{m-1}] \; , 
\label{mseries}
\end{equation}
where $a_{\nu}^0$ is given by Eq.~(\ref{anu0}).
The first order term is given trivially by $ E_{int}^{0(1)} = \lambda d N$. 

The second order term of the interaction energy is given by:

\begin{eqnarray}
E_{int}^{0(2)} & = & \lambda^2 d \sum_{i=1}^N \left[\sum_{j=1(\neq i)}^{2N} 
\frac{1}{2(j-i)} - \sum_{j=1(\neq i)}^{N} \frac{1}{j-i}\right] 
\nonumber \\& = &
\lambda^2 d \sum_{i=1}^N \left[\sum_{j=N+1}^{2N} \frac{1}{2(j-i)}\right] .
\label{ordersecond}
\end{eqnarray} 

This result can also be obtained by standard second order perturbation 
theory (\cite{DR00} and references therein). For large $N$ one obtains: 
$E_{int}^{0(2)} \approx \ln{2}\,\, \lambda ^2 d N = \ln{2} \, \, \lambda^2 
\omega_D$.

The calculation of the higher order terms is more difficult. We now make 
some approximations which enable us to find the $m$'th order within a factor 
of order unity. 

First we manipulate the $a_{\nu}^0$'s (\ref{anu0}) and obtain: 

\begin{equation}
a_{\nu}^0 = \frac{1}{2} \left[\sum_{k=1}^{\nu-1} \frac{1}{k} - 
\sum_{k=1}^{N-\nu} \frac{1}{k} + \sum_{k=N-\nu+1}^{2N-\nu} \frac{1}{k}\right] .
\label{anu0transformed}
\end{equation}
This can be approximated by: 

\begin{equation}
a_{\nu}^0 = \frac{1}{2}\ln[ 2(N+1)/ \nu - 1]- 
\ln[(N+1)/\nu - 1] \; .
\label{anu0approxi}
\end{equation}

For $\nu \ll N$ one obtains $a_{\nu}^0 \simeq - \ln{(N/\nu)}/2$, and for 
$\nu \simeq N$ 
(meaning $N-\nu \ll N$) one obtains $a_{\nu}^0 \simeq \ln{(N/\nu)}$. 

We now make a crude approximation:

\begin{equation}
\sum_{\nu=1}^{N}(a_{\nu}^0)^{m-1} \simeq 
(1-(\frac{1}{2})^{m-1}) \sum_{j=1}^{N}(\ln{(N/j)})^{m-1} \; .
\label{Nnu}
\end{equation}

This approximation is proper only for the $\nu$'s which are either small or 
close to $N$. However, since these $\nu$'s contribute the most to the sum, 
we expect this approximation to be correct within a numerical factor of 
order unity. 
Indeed, for $m=2$, the last sum in Eq. (\ref{Nnu}) approximately equals $N$, 
and the total result we obtain, $0.5 N$, is different than the 
correct result, $\ln{2} \cdot N$ only by a factor of order unity. 
Increasing $m$, the approximation becomes 
better, since the relative contribution of the levels far from $N$ decreases.
For $m>2$ we neglect the factor $(1/2)^m$ in (\ref{Nnu}). For $m \gg \ln{N}$ 
only the $j=1$ term in Eq. (\ref{Nnu}) contributes, and $E_{int}^{0(m)}$ is 
approximated by $\lambda^m \cdot \ln{N}^{(m-1)}$. For $m \ll \ln{N}$ one 
can show that 

\begin{equation} 
(m-1)! N - \ln{N}^{(m-1)} < \sum_{j=1}^{N}(\ln{(N/j)})^{m-1} < (m-1)! N 
\end{equation}
and therefore $\sum_{j=1}^{N}(\ln{(N/j)})^{m-1} \simeq (m-1)! N $.

Having thus established Eq.~(\ref{morder}), let us
now examine its consequences:

First, since the low powers in the series for $E^0_{int}$
fulfill the relation
$E_{int}^{0(m+1)}/E_{int}^{0(m)} \simeq  m \cdot \lambda$,
whereas the high  powers fulfill the relation
$E_{int}^{0(m+1)}/E_{int}^{0(m)} \simeq \lambda \cdot \ln{N}$,
 the series for $E_{int}^0$ 
does not have a single parameter describing the ratio between consecutive 
terms in the series.

Second,
while the high powers dictate the convergence radius of the series to be 
roughly $1/\ln{N}$, 
their contribution is larger than that of the low powers only for 
$\lambda \gtrsim 1/\ln{N} - 1/(\ln{N})^2$, introducing a scale of 
$1/(\ln{N})^2$ near $\lambda=1/\ln{N}$ (see also Sec.~\ref{secRichardson}). 
This can be seen by estimating the partial sum of Eq.~(\ref{mseries}) for 
$m \geq \ln{N}$ using the result in Eq.~\ref{limitm}. Taking $b'_m=1$ for 
all $m$ we get: 
\begin{equation}
\sum_{m=\ln{N}}^\infty \lambda^m (\ln{N})^{m-1} = 
\frac{(\lambda \ln{N})^{\ln{N}}}{\ln{N}(1-\lambda \ln{N})}\; \; .
\end{equation}
For $\lambda =  1/\ln{N} - c/(\ln{N})^2$, for any $c \gtrsim 1$, the above 
sum equals: 
\begin{equation}
\frac{(1-c/\ln{N})^{\ln{N}}}{c} \approx \frac{e^{-c}}{c} .
\end{equation}
which is smaller than one, while the low orders are 
proportional to $N$. 

Third, for $\lambda < 1/\ln{N} - 1/(\ln{N})^2$, where the low powers 
dominate, one readily obtains 
\begin{equation}
E_{int}^0 = \lambda N d + E^{0(2)}_{int} [1 + {\cal O}(1/\ln{N})] \, .
\label{eq:Eint0-result}
\end{equation} 
Here the first order term,
$\lambda N d$, is the 
Hartree term, and the
second order term, $E^{(2)}_{int}
 \approx \ln{2} \cdot \lambda^2 \omega_D$, 
is obtained from  Eq.~(\ref{ordersecond}). 
The order $1/\ln N$ can be
understood as follows. The third order term is smaller than the second 
order term by a factor of order $\lambda$, which is smaller than $1/\ln{N}$. 
All the higher orders are smaller than the second order term by a factor 
of order $1/(\ln{N})^2$ or smaller, and since there are about $\ln{N}$ 
such terms, their sum is also approximately $1/\ln{N}$ smaller than the 
second order term.

The corresponding perturbative result for the condensation energy,
$E^0_{cond} \equiv E^{pert}_{cond} $, now immediately follows by inserting
Eq.~(\ref{eq:Eint0-result}) into Eq.~(\ref{eq:definecond-energy})
[with $k=N$].
The result is given by Eq.~(\ref{eq:Econd-approximate}), namely 
$ E^{pert}_{cond} (\lambda) \simeq \ln{2} \cdot \lambda ^2 \omega_D \; $. 

\section{Series expansion of the { interaction} energy}
\label{appseries}

\begin{figure} 
        \begin{center}  
        \begin{picture}(1,0.618)
        \put(-0.05,-0.5){\psfig{figure=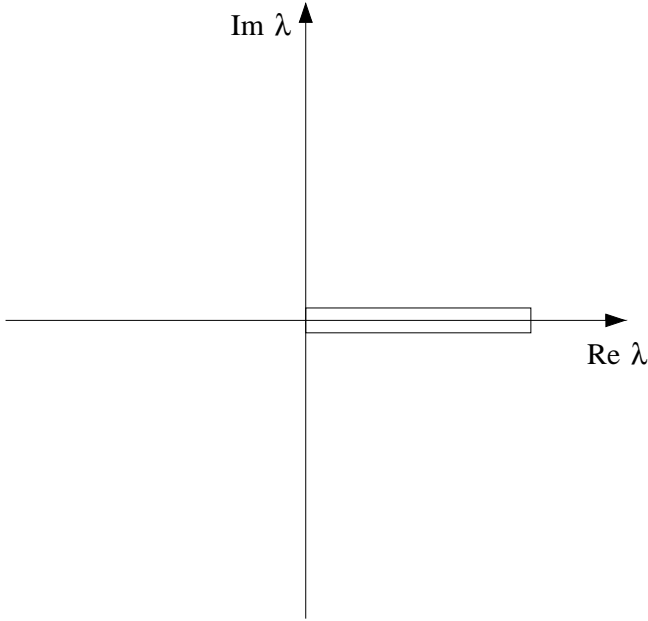,width=3.375in}}
        \end{picture} 
        \end{center}
        \vspace{3.5 cm}
        \caption{An integration contour in the complex $\lambda$ plane.}
        \vspace{0.8 cm}
\label{figcomplexcond}
\end{figure}

In Eq. (\ref{condmatveev}) we obtained an expression for the { interaction} 
energy, whose accuracy for different regimes in the range $\lambda<1/\ln{N}$ 
was obtained  in App.~\ref{appaccuracy}. We then expanded the 
result in a series in $\lambda$. This series converges, in the regime 
$\lambda<1/\hat{\ln}{N}$, to the approximate result. 
We were not able to find a series which converges to the exact result for 
the interaction energy in the above regime. 
However, we obtained results which 
suggest that the interaction energy is analytical on the positive real 
axis of the coupling parameter, 
and can be expanded in a series in $\lambda$ with a finite 
convergence radius which 
equals $1/(\hat{\ln}{N}+b)$, 
where $b$ is of order unity (Sec.~\ref{secRichardson}). 
As another check of the above statement 
we solved, numerically, Richardson's equations (\ref{Richeq}) for complex 
values of $\lambda$, and calculated the integral: 

\begin{equation}
\int_C E_{int}(z) dz ,
\label{complexcond}
\end{equation}
where $C$ is a contour circumventing the positive real axis (see 
Fig.~\ref{figcomplexcond}). 

The integral was calculated 
for various $N$'s in the range $4<N<64$ (a few contours for each $N$, 
each extending to a different value of $Re \lambda$, up to 
$Re \lambda = 0.7$). 
For all $N$ the integrals were zero 
within the numerical error, which suggests that there are no singularities 
on the positive real axis. 

The exact interaction 
energy is given by Eq.~(\ref{eq:Eint}). Expanding 
the exact expression 
as $E_{int} = 
d \sum_{\lambda = 1}^\infty \alpha_m \lambda^m $, 
we find that 
\begin{equation}
\alpha_m = \sum_\nu \left( (a_\nu^0)^{(m-1)} + 
\sum_{s=1}^{m-3} b_s (a_\nu^0)^s \right) .
\label{exactser}
\end{equation}

The approximation we make in Eq.~(\ref{condmatveev}) is equivalent to taking 
only the highest power ($m-1$) in Eq.~(\ref{exactser}) for each $m,\nu$, 
and neglecting the sum in the brackets.

\end{multicols} 

\begin{thebibliography}{10}
\bibitem{energies}
Here, and henceforth, we neglect factors of order unity when translating
  conditions and regimes from being stated in terms of $\lambda$ (which enters
  into the exponent) to being stated in terms of energies.
\bibitem{Pancake}
Orbital magnetization is usually neglected in ultrasmall
grains.~\cite{DR00} However, since we consider very large magnetic
fields, of order $\omega_D/\mu_B$, orbital effects might be important
unless the grains are pancake-shaped and the field is applied in
parallel to the flat direction, for the same reasons as apply to a
thin film in a parallel magnetic field.~\cite{MT94}
\end{thebibliography}
\end {document}